\documentclass[twocolumn,showpacs,preprintnumbers,amsmath,amssymb,unsortedaddress]{revtex4}

\usepackage{graphicx}  
\usepackage{dcolumn}   
\usepackage{bm}        

\newcommand{\beq}{\begin{equation}}
\newcommand{\eeq}{\end{equation}}
\newcommand{\beqa}{\begin{eqnarray}}
\newcommand{\eeqa}{\end{eqnarray}}

\def\bx{{\bf x}}
\def\br{{\bf r}}

\def\bk{{\bf k}}
\def\rhob{\rho_{\rm b}}

\def\del{\nabla}

\font\BF=cmmib10

\def\k{{\hbox{\BF k}}}
\def\x{{\hbox{\BF x}}}
\def\r{{\hbox{\BF r}}}

\def\q{{\hbox{\BF q}}}
\def\v{{\hbox{\BF v}}}
\def\u{{\hbox{\BF u}}}
\def\by{{\hbox{\BF y}}}

\def\tvk{{\hat{\k}}}
\def\tvq{{\hat{\q}}}

\def\fun#1#2{\lower3.6pt\vbox{\baselineskip0pt\lineskip.9pt
        \ialign{$\mathsurround=0pt#1\hfill##\hfil$\crcr#2\crcr\sim\crcr}}}

\def\Mpc{\, h^{-1} \, {\rm Mpc}}
\def\Gpc{\, h^{-1} \, {\rm Gpc}}

\def\lin{{\rm lin}}

\def\mc{{\rm mc}}

\def\MNRAS{{Mon.~ Not.~ R.~ Astron.~ Soc.~}}
\def\PR{{Phys.~ Rep.~ }}
\def\PRD{{Phys.~ Rev.~ D.~}}

\def\ApJ{{Astrophys.~ J.~}}

\def\ApJL{{Astrophys.~ J.~ Lett.~}}

\def\Nat{{Nature (London)~}}
\def\AstroPart{{Astro-particle Phys.~}}
\def\RvMP{{Rev.~ Mod.~ Phys.~}}

\begin{document}


\title{{\em Eppur Si Muove}:\\ 
  On The Motion of the Acoustic Peak in the Correlation Function}


\author{Robert E. Smith$^{1}$, Rom\'an Scoccimarro$^{2}$ and Ravi K. Sheth$^{1}$
\small \\ \vspace{0.2cm}
(1) University of Pennsylvania, 209 South 33rd Street, 
Philadelphia, PA 19104, USA.\\
(2) CCPP, Department of Physics, New York University, New York, NY 10003, USA. \\
email: {\tt res@astro.upenn.edu, rs123@nyu.edu, shethrk@physics.upenn.edu}}


\begin{abstract}
The baryonic acoustic signature in the large-scale clustering pattern
of galaxies has been detected in the two-point correlation
function. Its precise spatial scale has been forwarded as a rigid-rod
ruler test for the space-time geometry, and hence as a probe for
tracking the evolution of Dark Energy. Percent-level shifts in the
measured position can bias such a test and erode its power to
constrain cosmology. This paper addresses some of the systematic
effects that might induce shifts: namely non-linear corrections from
matter evolution, redshift space distortions and biasing. We tackle
these questions through analytic methods and through a large battery
of numerical simulations, with total volume of the order
$\sim100~[{\rm Gpc}^3\,h^{-3}]$.  A toy-model calculation shows that
if the non-linear corrections simply smooth the acoustic peak, then
this gives rise to an `apparent' shifting to smaller scales. However
if tilts in the broad band power spectrum are induced then this gives
rise to more pernicious `physical' shifts. Our numerical simulations
show evidence of both: in real space and at z=0, for the dark matter
we find percent level shifts; for haloes the shifts depend on halo
mass, with larger shifts being found for the most biased samples, up
to $3\%$. From our analysis we find that physical shifts are greater
than $\sim0.4\%$ at $z=0$. In redshift space these effects are
exacerbated, but at higher redshifts are alleviated. We develop an
analytical model to understand this, based on solutions to the pair
conservation equation using characteristic curves. When combined with
modeling of pairwise velocities the model reproduces the main trends
found in the data. The model may also help to unbias the acoustic
peak.
\end{abstract}


\maketitle


\section{Introduction}

Within the last few years the discipline of physical cosmology has
greatly benefited from a considerable influx of extremely high
fidelity data-sets, which have enabled measurements of the large scale
structure of the Universe to be made with unprecedented precision; and
together these data have led to the establishment of the `standard
model of cosmology': the flat, Dark Energy dominated collisionless
Cold Dark Matter (CDM) model
\cite{Spergeletal2006,Tegmarketal2006,Sanchezetal2006,Astieretal2006,Hoekstraetal2006,Coleetal2005,Eisensteinetal2005}.
Whilst the CDM particles are well founded from a particle physics
point of view, the Dark Energy may arise through a number of possible
mechanisms, most of which are of deep consequence to much of physics
if found to be true \cite{PeeblesRatra2003,DETF2006,ESO2006}. The task
of modern theoretical and observational cosmology, therefore, is to
construct robust tests to expose the true physical character of the
Dark Energy and hence differentiate between hypotheses.  A number of
experiments are currently underway with this sole purpose in view, and
many more are being planned for the future (see
\cite{DETF2006,ESO2006} and references there in for a comprehensive
review of current and future missions). The Dark Energy tests fall
into two main classes: those which perform geometric tests of gravity
and those which perform growth of structure tests. The geometric tests
are essentially the use of `standard candles' (Type Ia Supernova) and
`standard rods' (baryonic acoustic oscillations), whereas the growth
of structure tests, examine how the growth rate of perturbations
changes as a function of cosmological epoch.  Weak lensing by large
scale structure and the multiplicity function of clusters fall into
both categories and therefore potentially offer the most powerful
discriminatory means.  However, in order to make precise, accurate and
useful constraints on the Dark Energy, the systematics of each
experiment must be fully understood and controlled to sub-percent
accuracy \cite{DETF2006,ESO2006} -- the removal of `unknown unknowns'
is imperative.

For instance, the standard candle measurement from Type Ia supernovae
must address the issue of whether or not the ensemble of candles
evolves with redshift, i.e. through metalicity effects, or evolution
of the underlying host galaxy properties as a function of redshift.
Moreover until the `true' mechanism that drives the nova is
understood, it may be the case that this potential systematic can only
be quantified and eliminated once the data are in hand.

In this paper we shall restrict our attention to the second of the
geometric tests, that is the standard rod measurement from the
Baryonic Acoustic Oscillations (BAO). Like the standard candle test,
this method also suffers from potential systematics; the three knowns
in this case are: nonlinear mass evolution, non-linear bias and
redshift space distortions (hereafter, we shall refer to these
together as clustering systematics). However, unlike the case for Type
Ia Supernova, because the processes driving any possible evolution are
plausibly understandable {\em ab-initio}, there is not much room for
unknown unknowns and there is some hope for estimating and mitigating
these effects well-before the data streams in from the next generation
surveys. This is important because if the BAO peak is displaced by
even 1\%, this will induce a bias in the inferred value of the dark
energy parameter $w$ on the order of $~5\%$
\cite{Eisensteinetal2005,Anguloetal2007}.

The physical picture for the BAO signature is as follows: before the
epoch of recombination, acoustic oscillations were able to propagate
through the photon-baryon plasma at the sound speed, and these waves
were weakly coupled to dark matter through gravity. After
recombination the photons free stream out of the perturbations and
this gives rise to the observed CMB (\cite{Hinshawetal2006}), the dark
matter and segregated baryons then relax together over time and the
self-same acoustic features that are imprinted in the CMB become
imprinted in the dark matter distribution.  The characteristic scale
for the acoustic waves is set by the sound horizon at last scattering
$r_{*}$ (see \cite{EisensteinHu1998} for a description of how to
calculate this), and this in turn imprints a characteristic scale in
the pattern of galaxies and it is supposed that this has the
properties of a `standard rod'.

The BAO features have been detected by various groups: in the
two-point correlation function of Luminous Red Galaxies (LRG) by
\cite{Eisensteinetal2005}, and in the power spectrum of galaxies and
LRGs by
\cite{Percivaletal2001,Tegmarketal2004,Coleetal2005,Hutsi2006a,Hutsi2006b,Tegmarketal2006,Padmanabhanetal2006,Percivaletal2007a,Percivaletal2007b}.
The BAOs have also been the subject of much vigorous theoretical and
numerical research
\cite{Meiksinetal1999,BlakeGlazebrook2003,SeoEisenstein2005,White2005,Anguloetal2005,Springeletal2005,Dolneyetal2006,Bernstein2006,Huffetal2006,Eisensteinetal2006a,Eisensteinetal2006b,JeongKomatsu2006,Wang2006,Smithetal2007,Ma2007,Anguloetal2007,CrocceScoccimarro2007,MatarresePietroni2007}.
The question of whether there are non-linear effects at play on the
acoustic scale, is not an open question \cite{Meiksinetal1999},
however, whether these non-linearities give rise to an actual motion
of the acoustic peak -- apparent or physical -- is of great debate,
and the most recent literature concerned with this question reaches
conflicting conclusions: \cite{Guziketal2007} used the fitting formula
for the power spectrum from \cite{Smithetal2003} to conclude that,
there is a shift due to nonlinear mass evolution on the order of
$\sim2\%$ at $z=0$. \cite{SeoEisenstein2005} used numerical
simulations to show that there were changes to the broad band power
spectra of dark matter and haloes, and in both real and redshift
space, however they argued that provided these were accounted for, no
overall shift in the acoustic peak position would be induced.
\cite{Smithetal2007} used numerical simulations with improved
resolution to convincingly confirm the results from
\cite{SeoEisenstein2005}, that the power spectra were not immune to
strong broad band tilts. Based on these results they suggested that
percent level shifts in the position of the acoustic peak were highly
plausible. The main findings of these works were most recently
substantiated by \cite{Anguloetal2007}.  On the other hand,
\cite{Eisensteinetal2006a} used a model based on Lagrangian
displacements of the initial density distribution to argue that any
acoustic peak shift in the dark matter should be only of the order
$10^{-4}$ at $z=0$, although they do note that ``galaxy bias could
produce a sub-percent shift".  In addition, \cite{Ma2007} studied how
a relatively (by BAO standards) large peak in the initial power
spectrum evolved in numerical simulations and concluded that there
were no noticeable shifts, in agreement
with~\cite{Eisensteinetal2006a}.

In what follows, we examine this issue in detail. We do this in a
two-fold way: Firstly, we generate a large ensemble of large volume
numerical simulations to quantify the possible effects.  Secondly, we
develop a new analytical model, which is based on a new solution for
the pairwise conservation of particle pairs.  When combined with a
careful modeling of the divergence of pairwise velocities beyond
linear theory this method is shown to capture the main effects that
are found in the the simulations.

The subject of this paper is therefore to answer the following
important questions: Does non-linear evolution generate a displacement
of the peak of the correlation function? If so does the observed shift
depend on the halo/galaxy sample considered and how?  Recently, there
has been a number of different approaches to estimating the sound
horizon from observational data, however, so far as we know, it has
not been shown that any of these estimators are consistent, unbiased,
or indeed minimum variance estimators.  The results presented in this
work and from our previous study of the power spectrum
\cite{Smithetal2007,CrocceScoccimarro2007} should act as an important
empirical guide for constructing such quantities.

The paper is structured as follows: In Section \ref{getreal} we
discuss a toy model that shows that an effective smoothing of the
acoustic peak in the two-point function leads to an `apparent' motion
of the peak. Here we also show how if nonlinear evolution induces a
broad band tilt in the underlying linear power spectrum, further
shifts in the peak position are to be expected -- these we shall class
as `physical' shifts.  Then in Section~\ref{itmoves} we describe our
ensemble of numerical simulations and present our measurements for the
two-point correlation function of dark matter and haloes in real and
redshift space, including a detailed analysis of our data.  In
Section~\ref{andhow} we describe our new physical model and
demonstrate how it gives rise to a transformation of the structure of
the peak in the dark matter and halo correlation functions -- and that
this gives rise to a physical motion of the peak. We also compare our
analytic model with the results from the numerical simulation and show
that they are in close agreement. Finally, Section \ref{concl}
summarizes our results, and discusses them in the wider context.


\section{Apparent and physical shifts}
\label{getreal}

\subsection{Motivation}

Motivated by the calculation of the real space dark matter correlation
function in renormalized perturbation theory (hereafter, RPT)
\cite{CrocceScoccimarro2006a, CrocceScoccimarro2006b}, we can write
the observed correlation function in terms of the linear one through
the following relation:
\beq \xi_{\rm obs}(r) \approx \int \xi_{\lin}(r-r')\, K(r')\, d^3r' + \xi_{\rm
mc}(r)\ ,
\label{eq:xiRPT} 
\eeq
where the first term on the left-hand-side represents the linear
correlation function convolved with some symmetric kernel, $K(r)$, and
the second term, $\xi_{\rm mc}$, describes any effects due to
non-linear mode-coupling.  The distinction between these two terms may
be more clearly seen in Fourier space: the first term is directly
proportional to the linear power spectrum {\em at the same scale}, and
the second term represents a weighted sum over the information from
different neighboring wavemodes. Note that such decomposition can
always be made.  In RPT, the kernel $K$ is well approximated by a
Gaussian~\cite{CrocceScoccimarro2006b}, a result that becomes exact 
in the Zel'dovich approximation~\cite{B96,CrocceScoccimarro2006b,Eisensteinetal2006a}.

Setting aside $\xi_{\rm mc}$ for the moment, we remark that it is
sometimes thought that convolution with a Gaussian {\em does not} lead
to a shift in the BAO peak position.  In the following sub-section we
will show explicitly that this is not correct and that the convolution
with a symmetric filter {\em does shift the peak}, and that this is
solely due to the fact that $\xi_{\lin}$ is not symmetric about the
acoustic peak. However, as we mention in the following sub-section
this apparent shifting of the peak may be corrected for.

Returning now to the issue of mode coupling, as we will show in this
work through our numerical simulations and through our theoretical
analysis, the term $\xi_{\mc}$ in Eq.~(\ref{eq:xiRPT}) gives rise to
an actual `physical' shift towards smaller scales as the clustering
evolves.  For reasons which are now clear, we shall now refer to the
shifts that are generated by the first term as being apparent, and
those due to the second, as being physical.  In the next subsection we
present a toy-model to further illustrate the meaning of these terms.

 
\begin{figure*}
\begin{center}
{\includegraphics[width=0.6\textwidth]{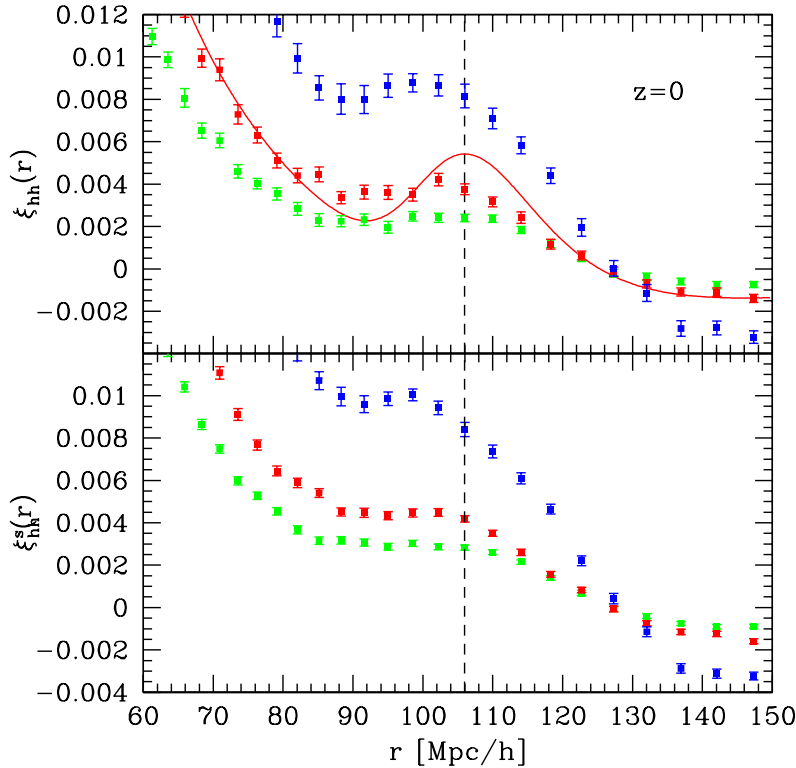}}
\caption{Halo correlation functions at $z=0$ in real (top) and
redshift (bottom) space.  Different symbols in each panel show results
for massive (top) to less massive halos (bottom).
Table~\protect\ref{halocat} gives the precise bins in halo mass.
Error bars come from the dispersion between the measured $\xi$ in our
50 simulations; a total volume of $105\,(\!\Gpc)^3$.  Solid line in
top panel shows the linear theory correlation function multiplied by
an arbitrary constant so that it approximately matches the signal from
the intermediate mass bin.  Vertical dashed line shows the position of
the acoustic peak in this linear correlation function: it lies at
$106\Mpc$.}
\label{xiz0}
\end{center}
\end{figure*}


\begin{figure*}
\begin{center}
{\includegraphics[width=0.6\textwidth]{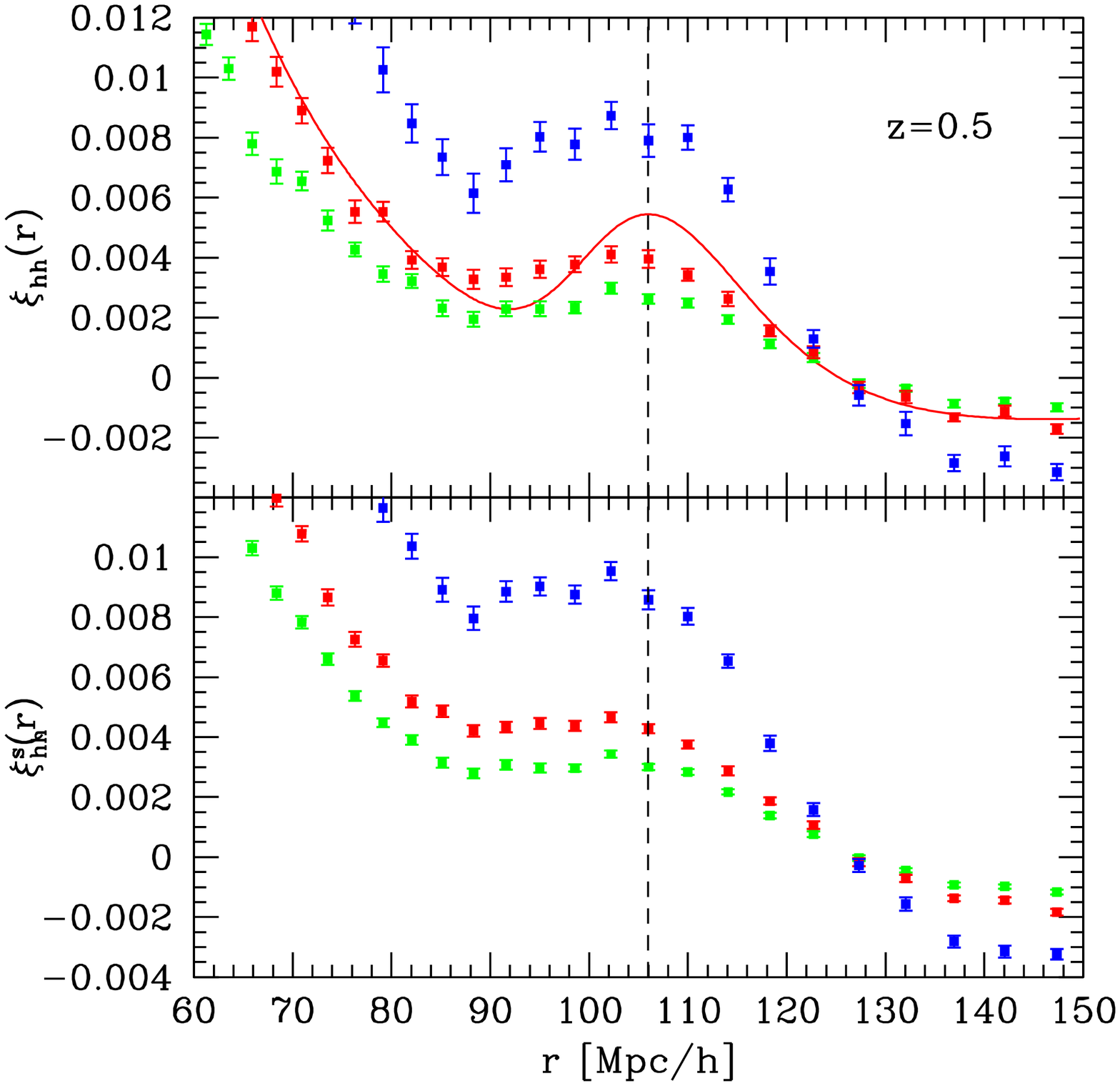}}
\caption{Same as Fig.~\ref{xiz0} but at $z=0.5$.}
\label{xiz0p5}
\end{center}
\end{figure*}


\begin{figure*}
\begin{center}
{\includegraphics[width=0.6\textwidth]{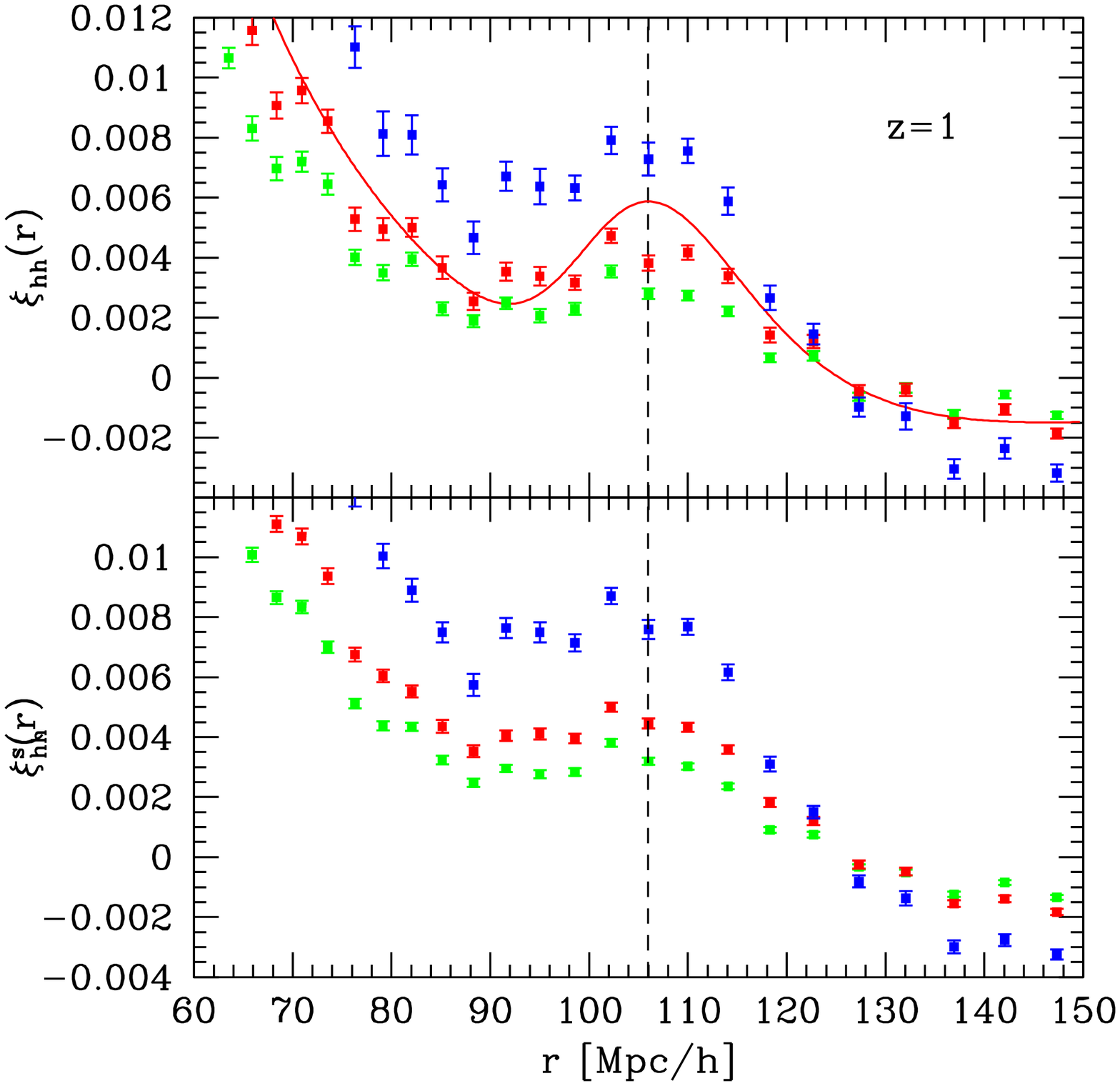}}
\caption{Same as Fig.~\ref{xiz0} but at $z=1$.}
\label{xiz1}
\end{center}
\end{figure*}


\subsection{A toy model for the shifts}
\label{itsnotreal}

Part of the following analysis was inspired by ideas first presented
by \cite{Guziketal2007}. In that work one of the issues addressed was
the apparent shift of the acoustic peak position, induced by an
inhomogeneous selection function. Here we use similar arguments, but
directly connected to the distortions induced by the non-linear
clustering transformation and bias, to examine the apparent shifts.
Those familiar with the analysis of \cite{Guziketal2007} may wish to
jump directly to Eq.~(\ref{eq:linpeak}), which should be familiar.

To begin our toy-model, let us suppose that the linear theory
correlation function can be well approximated by a power-law plus a
Gaussian bump with peak position located at $r_p$:
\beq
 \xi(r) = A_p\left(r_p\over r\right)^\gamma + 
A_G\, \exp\Bigl[-\frac{(r-r_p)^2}{2\sigma^2}\Bigr]\ .
\label{toymodel}\eeq
This is a reasonable approximation, since the transfer function can be
decomposed into a smooth component, which models the suppression of
dark matter fluctuations due to radiation dominated growth and baryon
drag effects, and an oscillatory piece that comes from the baryons
clumped around the sound horizon: i.e. $T(k)\equiv T_{\rm
smooth}(k)+T_{\rm BAO}(k)$ (see \cite{EisensteinHu1998}); on squaring
and Fourier transforming we get $\xi(r)\equiv \xi_{\rm
smooth}(r)+\xi_{\rm BAO}(r)$, where we have for simplicity neglected
the cross-terms from $T^2(k)$ (this is a toy-model). Restricting the
range of interest to be small enough so that $\xi_{\rm smooth}(r)$ is
close to a power-law, then we would have something like our
Eq.~(\ref{toymodel}).

The presence of the power-law means that the location of the local
maximum, say $r_{\rm max}$, will differ from $r_p$.  Requiring
$d\xi/dr = 0$ means
\beq A_p\,\gamma \left({r_p\over r_m}\right)^{\gamma + 1} = \left(1 -
 {r_m\over r_p}\right)\,{r_p^2\over \sigma^2}\,G(r_m) \ .  \eeq
If $r_m = r_p(1-\epsilon)$ then 
\beq
 A_p\,\gamma\,(1-\epsilon)^{-\gamma -1} = 
      \epsilon\,{r_p^2\over \sigma^2}\,G(r_m) \ .
\eeq
If $(r_p-r_m)^2 \equiv \epsilon^2 r_p^2\ll \sigma^2$ (meaning the
offset from $r_p$ is small compared to the width of the bump), then
this becomes 
\beq A_p\,\gamma\,(1-\epsilon)^{-\gamma -1} =
\epsilon\,{r_p^2\over \sigma^2}\, A_G\, \left[1 - {\epsilon^2
r_p^2\over 2\sigma^2}\right] \ .   \eeq 
To first order in $\epsilon$, this is
\beq \epsilon = \left[{A_G/A_p\over \gamma\,(\sigma^2/r_p^2)} -
 (1+\gamma) \right]^{-1}.  \eeq
The fact that we call it a bump means that $A_G>A_p$.  In addition, we
are interested in the case where $\sigma\ll r_p$, thus our final
expression for the peak in the linear correlation function is
\beq
 \epsilon_{\rm lin} \approx \gamma\,\left({\sigma\over r_p}\right)^2 \,
                          \left({A_p\over A_G}\right).
\label{eq:linpeak}
\eeq
This shows that the fractional shift from $r_p$ is large if $\gamma$
is large (meaning the amplitude of the power law component is changing
rapidly), or if $\sigma/r_p$ is large (meaning the bump is broad, so
the change in the amplitude of the power law component matters), or if
$A_p/A_G$ is large (meaning that the power-law component matters).

What concerns us now is: How does the peak scale change if one of our
clustering systematics alters one or all of these terms?  Suppose
$A_G\to A_G(1 + \delta_{A_G})$, $A_p\to A_p(1 + \delta_{A_p})$, etc.,
then we would have
\beq \epsilon \approx \epsilon_{\rm lin}
\left[\frac{(1+\delta_\gamma)(1+\delta_\sigma)^2(1+\delta_{A_p})}
{(1+\delta_{A_G})}\right]\ ; \eeq
and if these changes are all small, then $\epsilon\to \epsilon_{\rm
lin}(1 + \delta_{\epsilon})$
\beq \delta_\epsilon \approx \delta_\gamma + 2\delta_\sigma +
\delta_{A_p} - \delta_{A_G}\ .  \eeq
If the only effect of the clustering systematics is to smooth out the
spike to a bump, then they may simultaneously increase the width of
the peak and decrease $A_G$: i.e. $A_G\propto 1/\sigma$, implying that
$\delta_{\epsilon}\approx 3 \delta_{\sigma}$. However, because
$\delta_\sigma$ can be larger than $\sim0.1$, the effect on
$\epsilon\propto (1+\delta_\sigma)^3$ may be substantial.  We
emphasize that such an apparent shift would occur {\em even} if there
were no physical shift in the position of the peak.  Turning now to
the physical shifts: if $\delta_{\gamma}\ne0$ then we shall say that
our clustering systematics have changed the underlying power-law and
that this will lead to a physical motion of the acoustic peak.

Before we move on, we note that there are circumstances under which
the apparent shifts may be considered as benign and so removed, namely
the Gaussian smoothing case.  However, the physical shifts are more
pernicious and when these distortions are present it is not clear how
best to reconstruct the unperturbed peak for both of the shifts. We
shall reserve further discussion of this matter for our future
work. However, we note that for dark matter in real space these
effects can be calculated rather
precisely~\cite{CrocceScoccimarro2007}; in particular, the physical
shifts are more complicated than just an overall tilt of the
underlying power-law as described by our toy model.


\section{{Apparent and Physical} shifts from numerical simulations}
\label{itmoves}


\subsection{The ensemble of simulations}
\label{sims}

For the range of cosmologies that are acceptable, the BAO peak is
located at about $r_p\sim100\Mpc$.  A large simulation volume is
therefore required in order to minimize the cosmic variance in the
measurement on these scales and also to correctly account for the
mode-coupling from scales beyond $r_p$ that may drive evolution
\cite{Smithetal2007}. However, to control the sample variance down to
a level of a few percent requires the generation of a huge
computational volume. To make this task feasible, given our finite
computer resources, we decided to run a large ensemble of large
simulations as opposed to one single extremely large simulation.  As
we will show this allowed us to robustly answer the question as to
whether there is any apparent or physical evolution in the peak
position. These simulations will also allow us to assess how sensitive
future surveys will be to measuring the acoustic feature.  To this
end, we have run fifty realizations of cubic boxes with side $L_{\rm
box}=1280 \Mpc$, giving a total comoving volume of about
$105~(\!\Gpc)^3$, just under four times the volume of the Hubble
volume simulation. This is approximately the volume ADEPT plans to
survey, and is more than an order of magnitude larger than any current
or proposed ground based experiment \cite{Bennettetal2006}.

The cosmological parameters for the ensemble were selected to be in
broad agreement with the WMAP best fit model \cite{Spergeletal2006}:
$\Omega_m=0.27$, $\Omega_\Lambda=0.73$, $\Omega_b=0.046$, $h=0.72$ and
$\sigma_8(z=0)=0.9$.  For this cosmology, linear theory predicts the
position of the acoustic peak, i.e., the local maximum of the
auto-correlation function of dark matter, to occur at $106\Mpc$.

Each simulation was then run with $640^3$ particles. We used the {\tt
cmbfast} \cite{SeljakZaldarriaga1996} code to generate the linear
theory transfer function, and we adopted the standard parameter
choices, but took the transfer function output redshift to be at
$z=49$. The initial conditions for each simulation were then laid down
at $z=49$ using the 2LPT code described in
\cite{Scoccimarro1998,Crocceetal2006} and subsequent gravitational
evolution of the equations of motion was performed using the {\sf
Gadget2} code \cite{Springel2005}. Each realization ran to completion
in roughly 1900 timesteps from redshift $z=49$ to $z=0$, and the
comoving force softening was set at 70 kpc$/h$.

Haloes were identified in the redshift $z=0,\, 0.5$ and 1 outputs of
each realization, using the friends-of-friends algorithm with
linking-length parameter $l=0.2$ (this choice is standard).  Halo
masses were then corrected for the error introduced by discretization
of the halo density structure \cite{Warrenetal2005}.  Since the error
in the estimate of the halo mass diverges as the number of particles
sampling the density field decreases, we only study haloes containing
33 particles or more.  At each redshift we present results for the
three bins in halo mass.  These bins were chosen by counting down in
mass from the most massive halo, so that the number in each bin is the
same at each redshift.  Table~\ref{halocat} shows the resulting cuts
in halo mass, and the associated comoving number densities.


\def\nbarH{\bar{n}_H}


\begin{table}
\caption{\small Halo samples as a function of redshift.  Halos in the
 ``large'', ``intermediate'' and ``small'' mass bins $M>M_3$,
 $M_2<M<M_3$ and $M_1<M<M_2$, respectively.  Masses are in units of
 $h^{-1}M_{\odot}$ and comoving number densities $\nbarH$ in
 $(\Mpc)^{-3}$.}
\label{halocat}
\begin{ruledtabular}
  \begin{tabular}{lcccc}
& $z=0$ & $z=0.5$ & $z=1$ & $\nbarH$  \\ 
\hline 
\vspace{0.1cm}
$M_3$ & $ 1.5\times 10^{14}$ & $10^{14}$ & $ 5.7\times10^{13}$ 
    & $1.9\times10^{-5}$\\
\vspace{0.1cm}
$M_2$ & $ 7\times 10^{13}$ & $5\times 10^{13}$ & $ 3.1\times10^{13}$ 
    & $3.4\times10^{-5}$\\
\vspace{0.1cm}
$M_1$ & $ 4\times 10^{13}$ & $3\times 10^{13}$ & $ 2\times10^{13}$ 
    & $4.8\times10^{-5}$\\
\hline 
\end{tabular}
\end{ruledtabular}
\end{table}


\subsection{The measured correlation functions}
\label{thesignal}

The correlation functions were estimated using the standard estimator:
$\hat{\bar{\xi}}(r)=DD(r)/RR(r)-1$, where $\hat{\bar{\xi}}$ is the bin
averaged correlation function, $DD(r)$ is the number of true data
pairs in the bin and $RR(r)$ is the number of pairs expected after we
randomize the positions. We also note that when dealing with the
redshift space data, we apply the distortion separately in the $x-$,
$y-$ and $z-$directions and measure three correlation functions, these
are then averaged together to construct a single estimate for each
realization.

Figure~\ref{xiz0} shows the auto-correlation functions of the halos in
each of the selected mass bins in our $z=0$ outputs. Top and bottom
panels show $\xi(r)$ and $\xi^{\rm s}(r)$, the real and redshift space
correlation functions. We have chosen to show $\xi(r)$ rather than
$r^2\xi(r)$ because, as discussed earlier, the peak in the former is
more directly related to the sound horizon scale $r_s$.  The error
bars on the data points come from the scatter around the mean value of
$\xi$ as measured in the fifty realizations (i.e. from the diagonal
elements of the covariance matrix divided by the square-root of the
number of realizations, which for our case is: $\sqrt{50}\sim 7$).

The solid line in the top panel shows the dark matter correlation
function predicted by linear theory, multiplied by a constant factor
so that the curve approximately matches the signal seen in the
intermediate mass bin on scales $r\le 80\Mpc$.  The vertical dashed
line shows the location of the local maximum in this function:
$106\Mpc$.  Considering the results in real space (top panel), the
figure clearly shows that the local maxima of the measured correlation
functions are systematically shifted to smaller scales compared to
this mark. Moreover, it appears that the magnitude of the shift
steadily increases with halo mass. Turning to the results in redshift
space, we see that this effect is even more pronounced.  

Figures~\ref{xiz0p5} and~\ref{xiz1} show results at redshifts, $z=0.5$
and 1.  Although the distortions from the linear case appear to be
slightly smaller, we again see clearly that the trends are similar to
those of the redshift zero case. 

We now draw attention to another point of interest. As is expected,
these selected halo samples are significantly more clustered than the
mass.  The large-scale bias factors, as measured by the (square root
of the) ratio of the halo correlation function to that of the measured
dark matter on scales $\sim 70\Mpc$ (where nonlinear effects appear to
be small) are $b=1.4,~1.8,~2.6$ for the $z=0$ halos, $b=1.9,~2.3,~3.2$
for the $z=0.5$ halos, and $b=2.5,~3.0,~3.9$ for the $z=1$ halos --
with the most massive halos having the largest bias parameters.  What
is not so obvious now is that the halo clustering signal for each bin
at the three different redshifts is almost constant in time. For
reference, consider the linear theory growth factor which is smaller
by a factor of 0.785 between redshifts $z=0$ and $0.5$, so the
amplitude of $\xi_{\rm dm}$ drops between at $z=0$ and $0.5$ by a
factor of $\sim$0.615.  This result is a direct consequence of
studying the signal at fixed comoving number density: whilst the
clustering of the mass is much smaller at higher redshift, the high
redshift halos are significantly more biased.  At fixed number
density, the two effects approximately cancel out, keeping the net
clustering signal fixed. This is important in view of the fact that
galaxies of approximately constant comoving density represent a
popular choice for the target sample galaxy to measure the BAO
signature over a range of redshifts, i.e. the Luminous Red Galaxies
(LRG).


\subsection{Statistical properties and model fitting}

In this section we examine the statistical properties of the halo-halo
correlation functions and present our model fitting analysis.

\begin{figure*}
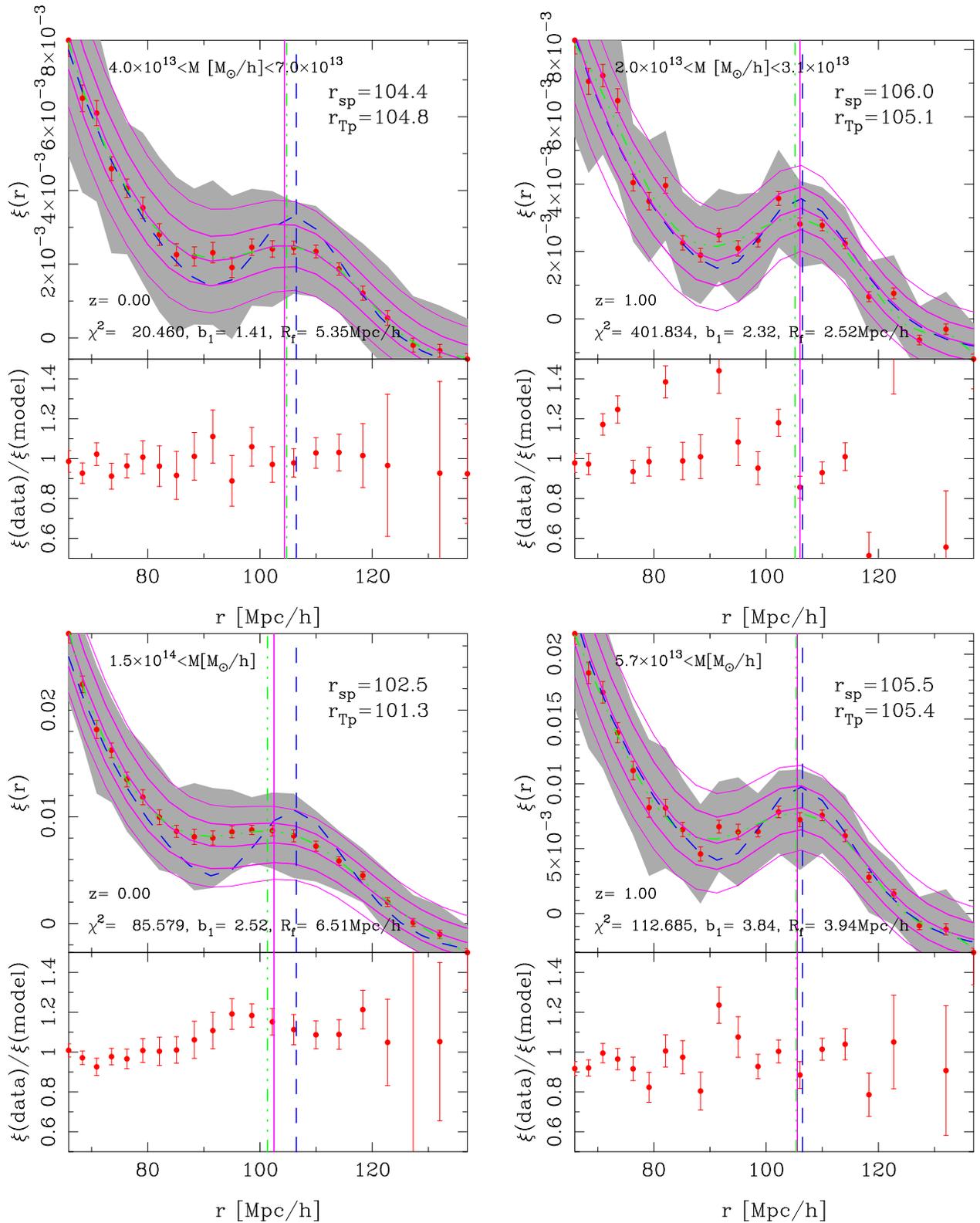

\centerline{
\includegraphics[width=8cm]{Fig.4a.eps}
\hspace{0.5cm}
\includegraphics[width=8cm]{Fig.4b.eps}}
\centerline{
\includegraphics[width=8cm]{Fig.4c.eps}
\hspace{0.5cm}
\includegraphics[width=8cm]{Fig.4d.eps}}
\caption{Mean (solid points), scatter (shaded region) and error on the
mean (error bars) for the halo-halo correlation functions measured in
the ensemble of 50 simulations. The long dashed curves show the
linearly biased, linear theory; the central solid curve shows linear
theory, smoothed with a Gaussian filter radius $R$ and linearly biased
$b$ (best fit values for these parameters are expressed in the figure
annotations). The inner and outer solid curves enclosing the best fit
model show the expected scatter in the continuum limit and the
discrete Poisson sample limit, respectively -- see text for full
explanation. The vertical lines represent the local maximum of the
linear theory $\xi$ (right most dash line) and the best fit smoothed
linear theory model (solid line) and the best-fit Tchebychev
polynomial fit (triple dot dashed lines).  The bottom panels show the
ratio of the measurements to the central solid line and again the
error bars are the errors on the mean.
\label{xirFit}}
\end{figure*}

\begin{figure*}
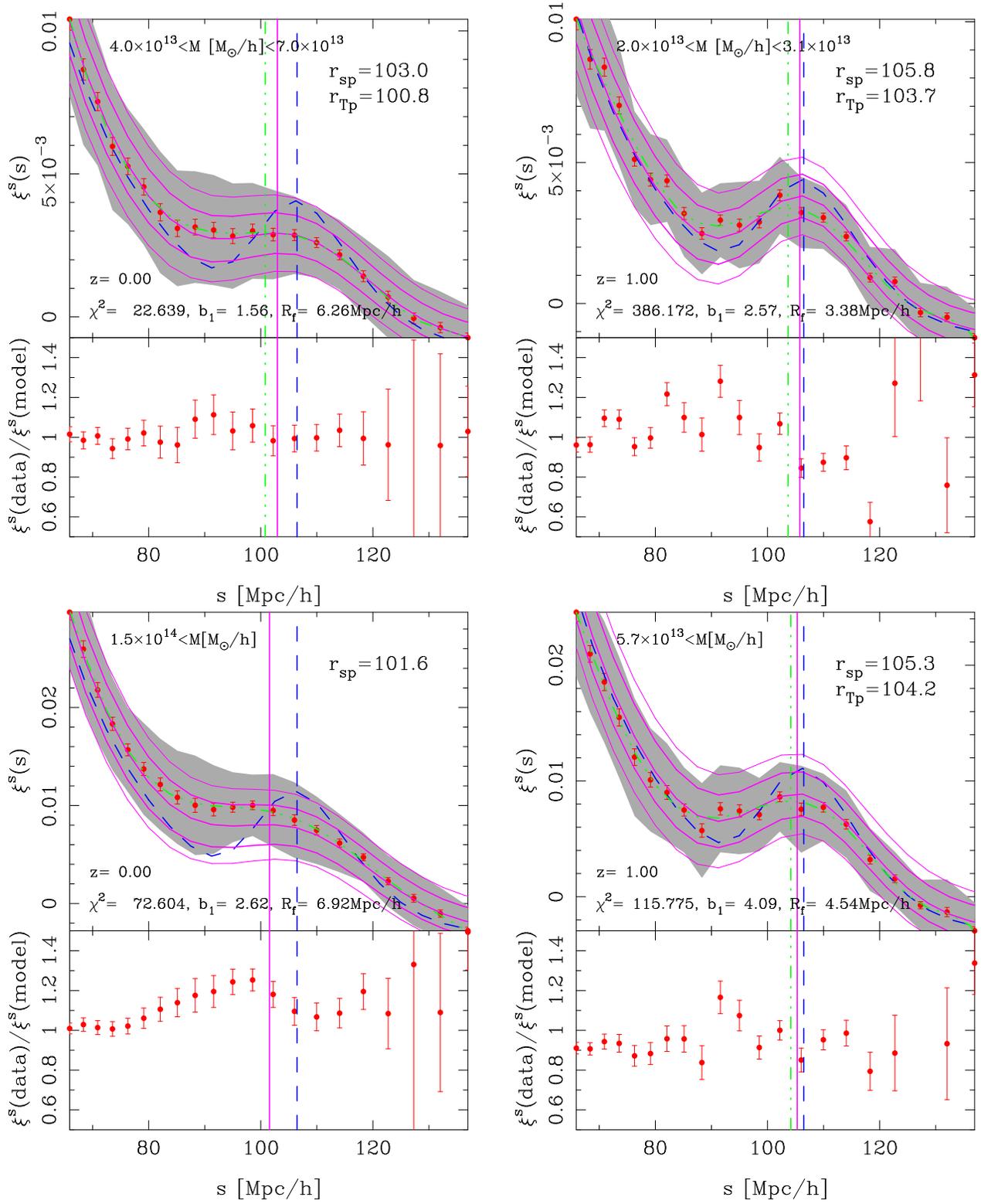

\centerline{
\includegraphics[width=8.cm]{Fig.5a.eps}
\hspace{0.5cm}
\includegraphics[width=8.cm]{Fig.5b.eps}}
\centerline{
\includegraphics[width=8.cm]{Fig.5c.eps}
\hspace{0.5cm}
\includegraphics[width=8.cm]{Fig.5d.eps}}
\caption{\small{Same as previous figure, but in redshift space.}
\label{xisFit}}
\end{figure*}

Figure~\ref{xirFit} shows again the simulation results in real space
and for the smallest (top) and largest (bottom) bins in halo mass, at
$z=0$ (left) and $z=1$ (right).  Figure~\ref{xisFit} shows again the
results in redshift space. In each panel, symbols show the mean value
of $\xi^{\rm hh}$ for the given bin in $r$, averaged over the 50
simulations; shaded regions show the standard deviation over the 50
realizations, and error bars show the error on the mean (they are
smaller than the shaded regions by a factor of $\sqrt{50}\approx 7$).

The first point to note is that the scatter amongst realizations is
remarkable, given that each one of our boxes is about three times
larger than the volume probed by the SDSS LRG sample. We further
emphasize that at least in redshift space this is likely to be a lower
bound on the true underlying scatter, since there is no contribution
to the variance from virial motions of the dark matter particles or
galaxies. Clearly, enormous volumes will be required to measure
$\xi^{\rm hh}$, and thus the galaxy correlation function, to the
precision required for percent precision cosmology, and this justifies
our earlier assertion at the beginning of this Section.

Comparing now the scatter exhibited in the $z=0$ real space low mass
halo sample with that found for the high mass sample, at a first
glance we see that the scatter appears to decrease as halo mass
increases; and this trend is also exhibited in redshift space data.
However as one goes to higher redshifts no obvious trend is apparent
between low and high mass samples. Comparing halo samples at the same
fixed number density but at different epochs, we see that the scatter
is much reduced for the low mass halo sample, but roughly constant for
the higher mass sample. This suggests that what is meant by `high' or
`low' mass is a very subjective quantity: `low' mass here must mean
relative to the typical halo mass at that epoch. However as we shall
show shortly these trends with halo mass and measured epoch can not be
characterized so na\"{i}vely.


We now turn to our modeling of the data. Based on our discussion from
Section \ref{getreal}, we now attempt to fit the correlation
functions by assuming that each can be described as a linearly biased
version of the linear theory correlation function, smoothed with a
Gaussian filter.  There are thus two free parameters for such
fits--the scale of the Gaussian smoothing filter $R_{\rm G}$ and the
overall amplitude of the bias, which we define $b_1^2\equiv {\xi}^{\rm
hh}(r)/\xi_{\rm Lin}(r)$, where ${\xi}^{\rm hh}(r)$ is the halo-halo
correlation function. Note that for this theoretical case we shall
assume that the transfer function and hence the cosmological model are
fully specified, which for our simulations they are of course, however
in reality one should consider fitting for these parameters jointly
with other cosmological parameters, albeit with constraints from
external data i.e. CMB etc., since these should not be considered
known {\em a priori} for a realistic case.

The best-fit model parameters for each sample were determined by
generating a 2D cubical grid of models over the ranges
$b_1\in\left[0.5,10.0\right]$ and $R_f\in\left[0.5,10.0\right]$,
spaced by steps of 0.01, and then computing the $\chi^2$ for each,
with the final best-fit model being identified as the one with the
minimum returned value. Explicitly the $\chi^2$ we minimize is
\beq \chi^2(\hat{b}_1,\hat{R}_f)= {\overline{\bf Y}}^T
 \hat{C}_{\left<\xi^{\rm hh}\right>}^{-1} {\overline{{\bf Y}}} \
 ,\label{chisquared}\eeq
where we define the ensemble average difference vector 
\beq {\overline{\bf Y}}\equiv {{\overline{\by}}-\by_{\rm
mod}(\bx|\hat{b}_1,\hat{R}_f)} \ ,\eeq
with ${\bar{\by}}=\left({\bar{\xi}}^{\rm hh}_1,\dots,{\bar{\xi}}^{\rm
hh}_N\right)$ and $\bx=\left(r_1,\dots,r_N\right)$, and where
$\by_{\rm mod}$ is a vector of model values. $\hat{C}_{\left<\xi^{\rm
hh}\right>}$ is our estimate of the covariance matrix of the mean
correlation functions
\beq \hat{C}_{\left<\xi^{\rm hh}\right>}\equiv 
\left[{\overline{\by^{T} \otimes \by}}
-{\overline{\by}}^T\otimes{\overline{\by}}\right]/N_{\rm real}\ ,\eeq
where $\otimes$ is the direct product operator and we divide by the
number of realizations, $1/N_{\rm real}$, because we are estimating
the covariances of mean quantities. The inversion of the above
Covariance matrix was performed using the SVD algorithm
\cite{Pressetal1992}. The models were fit over the range of scales
$(65\Mpc < r <140 \Mpc)$ in 21 equal bins in radius giving
$\sim3.5 \Mpc$ per bin.


Consider now the three thin solid curves, the central line represents
the best-fit smoothed and scaled linear theory correlation function;
and the best-fit values for $b_1$ and $R_f$, along with their
respective $\chi^2$ values are also reported in each panel.
Interestingly, we note that the bias factors in the redshift space
figures, recovered from this fitting procedure, roughly agree with our
earlier estimates from simply considering the data points around
$r\sim 70 {\rm Mpc}/h$.  For comparison the dashed lines show the
best-fit unsmoothed linear theory correlation function with a linear
bias. Clearly this model does not provide a reasonable fit to the
simulation data over the range of scales presented.

We may now make a prediction for the variance of the correlation
function in each bin by assuming that the underlying density field
from which the correlation functions were generated is well described
by a Gaussian random field. Following
\cite{Bernstein1994,Feldmanetal1994,Scoccimarroetal1999,MeiksinWhite1999,Cohn2006},
the covariance matrix for bin averaged correlation function, in the
limit of small bin sizes $\Delta r/r\ll 1$, can be written
\beqa \left[C_{\bar{\xi}^{\rm hh}}\right]_{ij} & \equiv &
\langle{\bar{\xi}}^{\rm hh}_i \bar{\xi}^{\rm hh}_j\rangle -
\langle\bar{\xi}^{\rm hh}_i\rangle\langle\bar{\xi}^{\rm hh}_j\rangle \nonumber \\ & = &
\frac{1}{V_{\mu}}\int {dk\, k^2\over 2\pi^2}\,j_0(kr_i)j_0(kr_j) \sigma^2_{P^{\rm hh}}(k)\ ,
\label{covar}
\eeqa
where $\bar{\xi}^{\rm hh}_i$ is the bin averaged correlation function
in bin $i$, $V_{\mu}$ is the simulation volume, $j_0= \sin(x)/x$ is
the zeroth order spherical Bessel function and where $\sigma^2_{P^{\rm
hh}}$ is the Gaussian variance {\em per mode} in the halo--halo power
spectrum, which for a discrete Poisson sampling of the halo field, can
be written
\beq \sigma^2_{P^{\rm hh}}(k) =
2\left[P^{\rm hh}(k)+\frac{1}{\nbarH} \right]^2
\label{eq:powvar} \ ;\eeq
where we write our smoothed halo-halo power spectrum, at linear order
in the over-density perturbation and bias, as
\beq P^{\rm hh}(k|R_f)\equiv V_{\mu}\langle\left|\delta^{\rm
h}(k|R_f)\right|^2\rangle = \hat{b}_1^2P_{\rm Lin}(k)W^2(k\hat{R}_f) \
. \label{eq:smoothlinearmodel}\eeq
For the purposes of numerical evaluation of the above formulae, we
follow \cite{Cohn2006} and note that one may rewrite the contribution
to the covariance coming from the term $1/\nbarH^2$ as follows:
\beq \frac{2}{V_{\mu}\nbarH^2}\int \frac{d^3\,k}{(2\pi)^3}
\,j_0(kr_i)\,j_0(kr_j) = \delta^{K}_{i,j}\frac{2}{\nbarH^2 V_{\mu}
V(r_i)}\ , \label{eq:shotnoise}\eeq
where the volume associated with each shell is: $V(r_i)=4\pi r_i^2
\Delta r$. The variance in the correlation function is then simply
given by setting $i=j$ in Eq.~(\ref{covar}).  Using our Gaussian
smoothed linear model for $P^{\rm hh}(k)$ ensures that the integrals
over the Bessel functions converge rapidly.  For a more detailed
discussion of convergence properties see \cite{Cohn2006}.

Considering again the best-fit smoothed model in each panel,
surrounding it are two sets of solid lines, a thick inner set and a
thin outer set.  The inner lines show the scatter between realizations
that one would predict using the continuum limit of
Eq.~(\ref{eq:powvar}), that is when $1/\nbarH\rightarrow 0$.  In this
case, the theoretical predictions clearly underestimate the true
scatter in $\xi^{\rm hh}$ for all bins in halo mass, with the
discrepancy being slightly worse for the lowest mass bin.  The outer
solid curves now show the effect of including the discreteness
contribution from $1/\nbarH$ in Eq.~(\ref{eq:powvar}). Note that in
implementing Eq.~(\ref{eq:shotnoise}) it was essential to correctly
account for the binning in the correlation function. In most cases,
and especially for the $z=1$ correlations, this additional
contribution provides a much improved description of the measured
scatter. However, at low redshift and for the lowest mass haloes
considered the theoretical estimate of the scatter appears too small.
This is likely a consequence of the growing non-Gaussian contributions
to the variance from the connected trispectrum and bispectrum,
\cite{Scoccimarroetal1999,MeiksinWhite1999,NeyrinkSzapudi2007} and
that the discrepancy between high and low-mass halo samples may be
caused by the effect of non-linear halo bias terms entering the
variance estimate. 

Returning to the issue of how the scatter depends on the
samples. Considering again our expression for the Gaussian error on
the power spectrum, Eq.~(\ref{eq:powvar}), we see that the relative
error per mode can be written:
\beq \frac{\sigma_P^{\rm hh}}{P^{\rm hh}}
=\sqrt{2}\left[ 1+\frac{1}{\nbarH b_1^2 P_{\rm Lin}(k,z)}\right] \ . \eeq
Thus we see that the relative scatter may increase in three ways: as
halo bias decreases; as halo number density decreases; and as the
power spectrum amplitude changes with time. For our constructed
samples these effects conspire in such a way that it is no longer
trivial to isolate trends. Rather than attempting to follow this path
we simply note that to say anything substantiative we must consider
the full covariance matrix, since a decrease in diagonal covariance
can be traded for an increase in off-diagonal covariance -- which is
just as important in the fitting. We shall reserve the important issue
of power spectrum and correlation function covariance for a future
study.

Lastly, we note that for very sparse samples of haloes the theoretical
prediction given by Eq.~(\ref{covar}) actually represents a lower
bound, since there should be a further (non-Gaussian) shot-noise
contribution of the order $\delta^{K}_{i,j} 2\bar{\xi}^{\rm
hh}_i/\left[\nbarH^2 V_{\mu} V(r_i)\right]$ \cite{Cohn2006}.  However,
for our purposes this term is unimportant, since $\bar{\xi}^{\rm
hh}_i\ll 1$.


\begin{table*}
\caption{\small Shifts in the BAO peak position as a function of halo
sample in real space.}
\label{tab:haloshiftreal}
\begin{ruledtabular}
  \begin{tabular}{cccccc}
SAMPLE & $R_f\ [\Mpc]$ & $b_1$ & $\delta_{\rm Sp} \
[\%]$ & $\delta_{\rm Tp} \ [\%]$ & $\delta_{\rm Phys}\ [\%]$ \\ 
\hline
& $z=0$ \hspace{0.3cm} $z=1$ & $z=0$ \hspace{0.3cm} $z=1$& $z=0$ \hspace{0.3cm} $z=1$& $z=0$ \hspace{0.3cm} $z=1$& $z=0$ \hspace{0.3cm} $z=1$ \\
\vspace{0.1cm}
$M_1$ & 5.35 \hspace{0.3cm} 2.52 & 1.41 \hspace{0.3cm} 2.32 & 1.5 
\hspace{0.3cm} 0.00  & 1.10 \hspace{0.3cm} 0.85 & 0.38 \hspace{0.3cm} 0.85\\
\vspace{0.1cm}
$M_2$ & 5.25 \hspace{0.3cm} 2.52 & 1.71 \hspace{0.3cm} 2.72 & 1.32
\hspace{0.3cm} 0.00  & 5.20 \hspace{0.3cm} 0.09 & 2.92 \hspace{0.3cm} 0.09\\
\vspace{0.1cm}
$M_3$ & 6.51 \hspace{0.3cm} 3.94 & 2.52 \hspace{0.3cm} 3.94 & 3.30 
\hspace{0.3cm} 0.47 & 4.40 \hspace{0.3cm} 0.57 & 1.13 \hspace{0.3cm} 0.09\\
\hline 
\end{tabular}
\end{ruledtabular}
\end{table*}


\begin{table*}
\caption{\small Shifts in the BAO peak position as a function of halo
sample in redshift space.}
\label{tab:haloshiftred}
\begin{ruledtabular}
  \begin{tabular}{cccccc}
SAMPLE & $R_f\ [\Mpc]$ & $b_1$ & $\delta_{\rm Sp} \
[\%]$ & $\delta_{\rm Tp} \ [\%]$ & $\delta_{\rm Phys}\ [\%]$ \\  
\hline
& $z=0$ \hspace{0.3cm} $z=1$ & $z=0$ \hspace{0.3cm} $z=1$& $z=0$ \hspace{0.3cm} $z=1$& $z=0$ \hspace{0.3cm} $z=1$& $z=0$ \hspace{0.3cm} $z=1$ \\
\vspace{0.1cm}
$M_1$ & 6.26 \hspace{0.3cm} 3.38 & 1.56 \hspace{0.3cm} 2.57 & 2.83
\hspace{0.3cm} 0.18 & 4.90\hspace{0.3cm} 2.16 & 2.07 \hspace{0.3cm} 1.98\\
\vspace{0.1cm}
$M_2$ & 6.97 \hspace{0.3cm} 4.04 & 1.91 \hspace{0.3cm} 2.88 & 4.60 
\hspace{0.3cm} 0.66 & 6.13\hspace{0.3cm} 1.20 & 1.51 \hspace{0.3cm} 0.57\\
\vspace{0.1cm}
$M_3$ & 6.92 \hspace{0.3cm} 4.54 & 2.62 \hspace{0.3cm} 4.09 & 4.40 
\hspace{0.3cm} 0.70 & NA  \hspace{0.3cm} 1.80 & NA   \hspace{0.3cm} 1.10\\
\end{tabular}
\end{ruledtabular}
\end{table*}


\subsection{Evidence for shifts}
\label{itsreal}

We now illustrate very clearly the effects of `apparent' shifts on the
peak position of the correlation function, arising from the operation
of smoothing, and explore the hypothesis that the peak position also
exhibits large-scale `physical' motion.

Before we commence with this investigation we note that since our
simulations used the transfer function generated at $z=49$, our linear
theory acoustic peak is slightly less sharp than that which obtains
from using the transfer function at $z=0$ (which is more
correct). This means that the apparent shifts will be overestimated,
since, as was discussed in Section \ref{getreal}, smoothing affects
more a weaker peak. For dark matter clustering, a calculation using
renormalized perturbation theory~\cite{CrocceScoccimarro2007} shows
that the apparent shift is overestimated by about a factor of two by
using the transfer function calculated at $z=49$. However, the
physical shifts are {\em not} changed significantly. 

Considering again Figs~\ref{xirFit} and \ref{xisFit}, as noted above,
the (blue) dashed lines in each panel show the associated biased but
unsmoothed linear theory correlation functions, and clearly these do
not provide a good fit to the data. The corresponding vertical dash
lines show the position of the local maximum of this curve -- the
unperturbed acoustic peak: $r_{\rm Up}=106\,h^{-1}\Mpc$. Now consider
the best fit smoothed models (central solid magenta), these clearly
(by eye) provide a much improved fit to the data. On finding the local
maximum of these smoothed models, we see that in all cases the peak
has been shifted to smaller scales. These are denoted in each panel by
the solid magenta vertical lines and with their values reported in the
top right of each panel as $r_{\rm Sp}$, with `Sp' meaning smoothed
peak.

Considering these smoothed model inferred peak positions, we note
several important trends:
\begin{itemize}
\item All maxima lie on smaller scales than $r_{\rm Up}$;
\item For halo samples considered at the same epoch and in both real
and redshift space, the shifts from $r_{\rm Up}$ increase with
increasing halo mass;
\item Considering halo samples of the same fixed number density at
different redshifts, the shifts are reduced for the higher redshift
samples;
\item Shifts are increased in the redshift space;
\item Best fit filter scale increases with halo mass.
\end{itemize}
The spread of values in the smoothed model shifts $\delta_{\rm
Sp}\equiv {\left[r_{\rm Up}-r_{\rm Sp}\right]}/r_{\rm Up}$, span the
range $\delta_{\rm Sp} \sim 1.0\%-5\%$, with the largest values being
obtained as per the trends described above.  These shifts, it can be
argued \cite{Eisensteinetal2006a}, fall under the banner of apparent
shifts -- arising from the local collapse and rearrangement of matter.
However, following our discussion of the transfer function, we expect
that these shifts would be reduced by a factor of $\sim2$ for the
$z=0$ transfer function: $\delta_{\rm Sp} \sim 0.5\%-2.5\%$. We also
note that in the recent literature a number of procedures have been
proposed to tackle these apparent shifts and, modulo the choice of
filter function, most of these methods should be able to successfully
correct for these effects. However, we now draw attention to the last
of our bullet points and note the fact that the best fit filter scale
$R_{f}$, increased with halo mass. This implies that methods that are
tested and tuned to extract BAO information using only the dark matter 
distribution will fail to incorporate this effect -- we return to this 
in Sec.~\ref{ssec:alternate}. 

Turning now to the question of `physical' shifts, it is clear that the
smoothed model does not provide consistently good fits to the
measurements for all our samples. To see this more clearly, the bottom
section of each panel shows the ratio of the measured points to the
best-fit smoothed linear model.  From examination of these results it
is clear that there is some evidence for structure in these residuals
--- typically, on scales smaller than the true acoustic peak position
we find that the data points lie above the best fit smoothed
model. This trend is most apparent for the present day high mass halo
samples, but is less clear for the lowest mass. This can be further
quantified by use of the $\chi^2$ test as an indicator for the
`goodness-of-fit': for $N=21$ data bins and $m=2$ parameters, the
probability $P(\chi^2>36.19|n=19)=0.01$. Thus on inspection of the
$\chi^2$ values in Figs~\ref{xirFit} and \ref{xisFit} we see that only
in two instances are the mean data in agreement with this and these
are for the present day low mass samples in real and redshift
space. Based on these data we are led to reject our null hypothesis
and accept the possibility that there {\emph is} a physical motion of
the peak.

One alternative to the `physical motion' hypothesis is that the filter
choice is somehow special -- and had we chosen the `special' filter
then this would reconcile our results. This view is problematic, since
in order to model all of the above trends such a filter would have to
be highly contrived. Thus, based on the above evidence, it seems that
something like the second term in Eq.~(\ref{eq:xiRPT}) is present and
generating a shift in the position of the peak. 

In the next section, we provide details of a physically motivated
model that may offer some insights into the origin of the dependence
on halo mass of the shifts in the acoustic peak position.

Before continuing, it is of interest to characterize the peaks in the
correlation function data using a purely artificial parameterized model
that simply matches the data in the least square sense. For this we
write the model of the correlation function as a sum over the
orthogonal Tchebyshev polynomials, $T_i(x)$, i.e.
\beq y_{\rm mod}(r)=\xi^{\rm hh}(r)=\sum_{i=0}^{m}a_i T_i[x(r)] \ ;
\eeq
and $x(r)$ here is a mapping that transforms the $r$-range into the
range $x\in\left[-1,1\right]$ so that we may use the normalized
polynomials. We then, as before, find the coefficients $a_i$ that
minimize our $\chi^2$ Eq.~(\ref{chisquared}) and polynomials up to
degree 9 were used to describe the data and, for simplicity, in the
fitting we have used only the diagonal elements of the covariance
matrix to make the constraints. These artificial models along with the
locations of their local maxima at the acoustic scale are presented in
Figs~\ref{xirFit} and \ref{xisFit} as the (green) triple-dot dash
curves, and the values of the maxima are noted in the top-right hand
corner of each panel as $r_{\rm Tp}$. For the high mass samples the
shifts, $\delta_{\rm Tp}\equiv[r_{\rm Up}-r_{\rm Tp}]/r_{\rm Up}$,
are significantly enhanced relative to linear model, and for the case
of the highest mass halo sample at $z=0$ in redshift space, the best
fit model had no local maximum.

We may now be more definite in what we mean by a physical shift: we
shall say the percentage physical shift away from the true peak is
$\delta_{\rm Phys}\equiv \left| \delta_{\rm Sp}-\delta_{\rm
Tp}\right|$. The physical shifts appear to be smaller than the
apparent shifts and are roughly of the order $\sim0.4-3.0\%$ at
$z=0.0$ for real and redshift space data and they are somewhat
mitigated at higher redshifts.  Tables \ref{tab:haloshiftreal} and
\ref{tab:haloshiftred} collect together the apparent, total and
physical shift values for the halo samples in real and redshift space,
respectively.  As we shall discuss in the following section these
physical shifts are not accounted for in the recent methods proposed
to correct for the non-linear evolution of the BAO peak position.


\begin{figure}
{\includegraphics[width=0.45\textwidth]{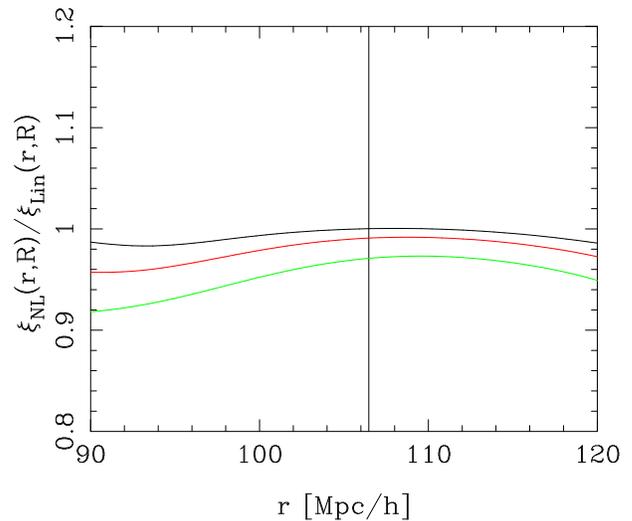}}
\caption{Expected behavior of the residuals in the correlation
function arising from the alternative approach to BAO model fitting of
\cite{Eisensteinetal2006a}, where the broad-band, but smooth, linear
power is restored. From top to bottom the lines show the effects where
a smoothing scale of $R_f=\{2, 4, 8\} \Mpc/h$ were taken. This method
does not replicate the $S$ shaped structure of the residuals found in
the data.}
\label{fig:AlternateModel}
\end{figure}


\subsection{Alternative BAO reconstruction methods}\label{ssec:alternate}

Recently a number of procedures have been proposed to correct the BAO
peak position for the effects of the large scale structure systematics
(see for example
\cite{SeoEisenstein2005,Eisensteinetal2006a,Huffetal2006,Percivaletal2007a,Anguloetal2007}).
These methods essentially allow for smoothing in configuration space
and possible broad band tilts in the underlying power spectrum. Since
the correlation function is the Fourier space dual of the power
spectrum, these methods should also be equally applicable in
configuration space.

Firstly, consider those methods that attempt to correct the measured
power spectrum for tilting by fitting away an arbitrary constant --
this is in response to ideas from the Halo Model that lead us to
consider a generalized Poisson shot-noise correction (see for example
\cite{SchulzWhite2006}). On Fourier transforming this model it does
nothing to the peak of the correlation function and so may be
dismissed.

Secondly, consider a model that is designed to damp out the acoustic
oscillations, but then restore the linear theory power using a smooth
(no BAO) version of the linear power \cite{Eisensteinetal2006a}, e.g.
\beq P_{\rm NL}(k|R)={\rm e}^{-k^2R^2}\,b^2 P_{\rm Lin}(k) + (1-{\rm
e}^{-k^2R^2})\, b^2 P_{\rm Lin}^{\rm smooth}(k),
\label{PtiltCorr}
\eeq 
In the configuration space the first term transforms into the smoothed
linear model Eq.~(\ref{eq:smoothlinearmodel}), and as we saw this will
generate apparent shifts. Considering the second term, we see that
this function has no information about the acoustic scale.  
Fig.~\ref{fig:AlternateModel} shows that this model for a range of 
smoothing filter scales - in all cases, it is flat around 
the acoustic peak. Moreover, the ratio
 $\xi_{\rm NL}(r|R)/\xi_{\rm Lin}(r|R)<1$ for all smoothing scales, 
whereas the measured residuals can exceed unity on scales smaller 
than the acoustic peak (cf. Figures~\ref{xirFit} and~\ref{xisFit}).  
We therefore deduce that a model like equation~(\ref{PtiltCorr}) 
is inadequate.

Lastly, we note that a more sophisticated method for correcting the
spectrum for the non-linear systematics was proposed by
\cite{Percivaletal2007a}. However, this method was recently looked at
in great detail by \cite{CrocceScoccimarro2007}, for the most 
optimistic case - dark matter in real space.  
There it was shown that, although the method accounts for
broadband tilting of the underlying power spectrum, it was unable to
correct for the shift of the peak due to mode-coupling. Since these
mode coupling terms are even more enhanced in the halo-halo power
spectrum (see \cite{Smithetal2007}) we expect that this method will 
not correct for all of the physical shifts found in the previous 
section.

In passing, we note that it is not straightforward to draw a 
direct connection between what we call the physical shift and what
\cite{CrocceScoccimarro2007} describe as mode-coupling shifts. 
It is likely that what we have called a physical shift is an
underestimate of the mode-coupling effects. To understand why, 
we note that whilst the Renormalized Perturbation Theory formalism 
has yet to be extended to Haloes, we may suppose that the RPT 
decomposition of the 2-pt clustering signal will still be valid,
i.e. there is some propagator which multiplies the linear theory 
power, and some sum of mode-coupling terms.  
Suppose the halo propagator has almost the same form as the dark 
matter propagator, so the effects of the non-linear bias mainly 
affect the mode coupling pieces.  Since the propagator is akin to 
a Gaussian smoothing term, this term acts just like the simple 
Gaussian smoothing model we fit to the simulation data.  
If the effects of the other (mode-coupling) term were negligible, 
or did not depend strongly on halo mass, then we would expect the 
scale of the best fit smoothing filter to also be independent of 
halo mass.  It is not, suggeting that the mode-coupling term depends 
on halo mass.  That the scale of the best-fit smoothing filter is 
larger for the more massive halos 
(Table~\ref{tab:haloshiftreal} and~\ref{tab:haloshiftred}) indicates 
that our best-fitting Gaussian filter is trying to account for some 
of the shifting that is actually due to the mode-coupling terms.  

We shall now pursue an analytic approach that we think provides
insight in to the physics behind the shifts.


\section{A physical model for the shifts}
\label{andhow}

This section presents a simple physical model for estimating the
effects of nonlinear clustering and bias on the position of the local
maximum of the correlation function.  \cite{CrocceScoccimarro2007}
discuss a more accurate model for the correlation function of the dark
matter; the approach below allows one to address how the peak shifts
are affected if the measured correlation function comes from a biased
tracer of the dark matter field.


\subsection{The pair conservation equation}

The perturbed continuity equation for the collisionless CDM fluid can
be written, 
\beq
 \frac{ \partial \left[1+\delta(\bx,\tau)\right]}{ \partial \tau } + 
 \nabla \cdot \Big[ (1+\delta(\bx,\tau))\, \v(\bx,\tau) \Big] = 0.
\label{cont}
\eeq
where 
$\delta(\bx,\tau) \equiv [\rho(\bx,\tau)-\rhob(\tau)]/\rhob(\tau) \ ,$
is the dimensionless density perturbation at comoving position $\bx$
and conformal time $\tau$ ($d\tau\equiv dt/a(t)$, where $a(\tau)$ is
the expansion factor from the Friedmann equation); $\rhob(\tau)$ is
the homogeneous background density; and ${\v}(\bx,\tau)\equiv
\bx^{\prime}\equiv d\bx/d\tau$ is the proper peculiar velocity field
\cite{Bernardeauetal2002,Peebles1980}.

We can now use Eq.~(\ref{cont}) at position 1, say, multiply by $(1+\delta_2)$ for
position 2, and add the same expression with indices $1$ and $2$
interchanged $[\delta_i\equiv\delta(\bx_i)]$.  Taking expectation
values of the result yields the pair conservation equation
\cite{Peebles1980,NityanandaPadmanabhan1994,Fisher1995,Juszkewiczetal1999}:
\beq \frac{ \partial [1+\xi(\br,\tau)] }{ \partial \tau } + 
\nabla \cdot \Big[ [1+\xi(\br,\tau)]\, \v_{12}(\br,\tau) \Big] = 0,
\label{paircon}
\eeq
where the divergence is with respect to the vector that separates 
the pair $\r=\x_1-\x_2$, and the pairwise infall velocity is 
\beq
\v_{12}(\br,\tau) \equiv \frac{ \langle (1+\delta_1)(1+\delta_2)(\v_1-\v_2) 
\rangle }
               { [1 + \xi(\br,\tau)] } ; 
\label{v12}
\eeq 
where by statistical isotropy  we used that 
$\left<\delta_1\v_1\right>= \left<\delta_2\v_2\right>=0$.

We can rewrite Eq.~(\ref{paircon}) in a more convenient form, by
changing time variable from conformal time $\tau$ to the linear growth
factor $D_+$.  In particular, if
\beq \eta \equiv \ln D_+ , \label{timevar}
\eeq
then $d\tau = d\eta/ ({\cal H} f)$, where ${\cal H} = d\ln a/ d\tau$
and $f = d \ln D_+/d\ln a$.  We may also write velocities in a similar
fashion and scale out their dependence on linear theory.  Namely, $\v
= - {\cal H} f \u$, where $\nabla \cdot \u = \delta$ in the linear
theory.  Then, dividing Eq.~(\ref{paircon}) by $[1+\xi(\br,\tau)]$
yields,
\beq \frac{\partial \ln [1+\xi(r,\eta)] }{\partial \eta } - 
\u_{12} \cdot \nabla \ln [1+\xi(r,\eta)] = \nabla \cdot \u_{12}(\br,\eta)
\label{paircon2} \ .
\eeq

Owing to the fact that large-scale flows have no vorticity, the
pairwise velocities are directed along the separation unit vector
$\hat{\br}$, so $\u_{12}= u_{12}\, \hat{\br}$. Hence
Eq.~(\ref{paircon2}) becomes,
\beq
 \frac{ \partial  \ln [1+\xi(r,\eta)] }{ \partial \eta } 
 -  u_{12}(r,\eta) \frac{\partial \ln [1+\xi(r,\eta)])}{\partial r} 
 = \Theta(r,\eta)\ ,
\label{paircon3}
\eeq
where we have defined $\Theta(r,\eta)\equiv \nabla \cdot
[u_{12}(r,\eta){\hat\br}]$ to be the divergence of the pairwise infall
velocities $u_{12}(r)$. Note, that this equation may be thought of as
a differential equation for $\ln (1+\xi)$ given an ansatz for $u_{12}$
\cite{NityanandaPadmanabhan1994}, or `vice-versa'
\cite{Juszkewiczetal1999}.


\subsection{Solution by characteristics}

The general solution of Eq.~(\ref{paircon3}) can be found by the
method of characteristics (see for example
\cite{LandauLifschitzFluids}), which illustrates quite clearly how any
feature in the correlation function will move as clustering develops.

The continuity equation (and thus the pair conservation equation) is a
prime example of a hyperbolic partial differential equation.
Information propagates from the initial conditions to the final
conditions through curves, called characteristics. The characteristics
are simply the equations of motion of pairs,
\beq
\frac{ d r}{d \eta} = -u_{12}(r,\eta).
\label{charac}
\eeq
The solution of this equation gives $r(\eta)$, and this converts the
left hand side of Eq.~(\ref{paircon3}) into a total derivative.  Thus,
one obtains an ordinary differential equation along the
characteristics:
\beq
\frac{ d \ln [1+\xi(r,\eta)] }{ d \eta } = 
\Theta(r,\eta) \ , \label{alongC}
\eeq
and it should be understood that it is a function of time $\eta$ only,
after using the characteristic solution $r(\eta)$,
Eq.~(\ref{charac}). Thus Eq.~(\ref{alongC}) simply gives the
logarithmic rate of change of the two-point correlation function as it
evolves along the characteristic curve. The fully evolved correlation
function may then be obtained straightforwardly, at any chosen epoch,
through integration along the characteristic between the initial and
final epoch:
\beqa 1+\xi(r, \eta) & = &\Big( 1+\xi_0[r_0(r,\eta)] \Big) \nonumber \\ 
& & \times \exp \Big[ \int_0^\eta \Theta[r_{\eta'}(r,\eta),\eta']
\, d\eta' \Big],
\label{formsol}
\eeqa 
where $r_0(r,\eta)$ is the initial separation that corresponds to $r$
at time $\eta$, and similarly for $r_{\eta'}$. The exponential factor
comes from the fact that the correlation function is not conserved
along characteristics because the right hand side of
Eq.~(\ref{alongC}) is non-zero.  Since we are mostly interested in
significant growth after the initial perturbations are laid down
($\eta \gg \eta_0$), the term in the first parenthesis can be safely
approximated as unity.  Hence, all the evolution is encoded in
$\Theta$ and the characteristics.  Note that this solution is {\em
exact}; it only becomes useful, though, if one can model the pairwise
infall velocities.


\begin{figure}
\begin{center}
{\includegraphics[width=0.5\textwidth]{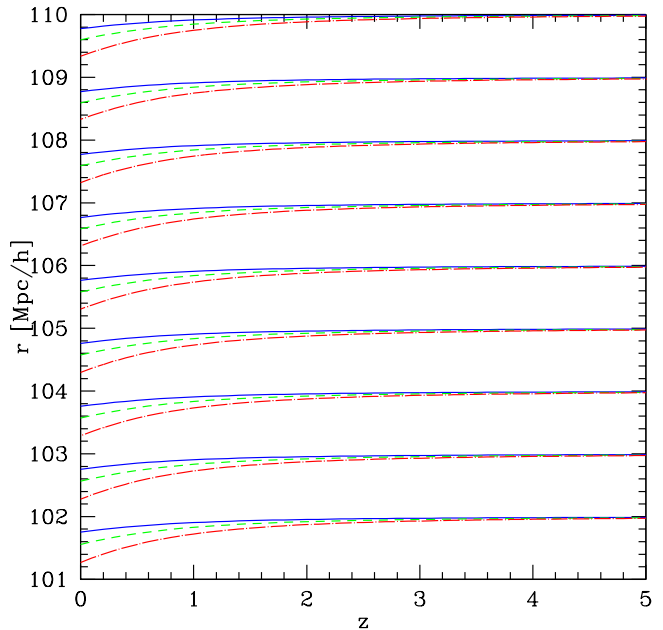}}
\caption{The flow of characteristics in linear theory for initial
separations close to the acoustic peak of the two point function,
every 1 $\Mpc$.  For each scale we show results for dark matter
(solid, solutions of Eq.~\ref{prop}), and linearly biased tracers
(solutions of Eq.~\ref{propbias}) having $z=0$ bias factors of $b=1.4$
(dashed) and $b=2$ (dot-dashed).  The peak in the linear correlation
function is located at $r=106 \Mpc$ for the cosmological model we use
in this paper.  }
\label{characFig}
\end{center}
\end{figure}


\subsection{Linear theory velocities}

For what follows, it will be convenient to define 
\beq \bar\xi_0(r_0,\eta_0) \equiv e^{2\eta_0}\, \frac{3}{r}  
\int \frac{P_0(k)}{k}\, j_1(kr)\,
 d^3k, \eeq
where $P_0(k)$ is the power spectrum at some initial time
$\eta_0\equiv 0$ and where $j_1(y)\equiv [\sin(y) -y \cos(y)]/y^2$ is
the first order spherical Bessel function.  In linear theory, pairwise
infall velocities, at time $\eta$, can be written
\cite{Peebles1980,Fisher1995}
\beq
u_{12}(r,\eta) = 2\, {\rm e}^{2\eta}\,\int
\frac{P_0(k)}{k}\, j_1(kr)\,d^3k
= \frac{2r}{3}\, {\rm e}^{2\eta}\, \bar\xi_0(r)\ .
\label{u12lin}
\eeq
\cite{velocityfield}.  The divergence of pairwise velocities in linear
theory can be obtained directly from Eq.~(\ref{u12lin}) by taking the
divergence,
\beqa
\Theta(r,\eta) & = & \nabla_r \cdot [u_{12}(r)\hat{\br}] 
\equiv \frac{1}{r^2}\frac{\partial}{\partial r}[r^2 u_{12}(r)] \ ,\nonumber\\ 
& = & 2 \,{\rm e}^{2\eta}\,\int P_0(k)\,j_0(kr)\,d^3k \ ,\nonumber \\
& = & 2\, {\rm e}^{2\eta}\, \xi_0(r)\ , 
\label{thetalin}
\eeqa
with $\xi_0$ the initial (linear) correlation function at $\eta_0=0$.

Eq.~(\ref{u12lin}) allows us to solve for the characteristics in
linear theory. Two-point information at separation $r_0$ and time
$\eta_0=0$ propagates by time $\eta$ to a separation $r$ (less than
$r_0$, due to clustering) so that, from Eq.~(\ref{charac})
\beq
 {\rm e}^{2\eta}-1 
   = \int_{r}^{r_0} \, \frac{3}{\bar\xi_0(r)}\frac{dr}{r}\ .
\label{prop}
\eeq

Figure~\ref{characFig} shows the solution of this equation ($r$ as a
function of redshift) for initial separations $r_0$ close to the
acoustic peak of the two-point correlation function. If this were the
only effect, i.e. if the right hand side of Eq.~(\ref{alongC}) were
zero, then the correlation function would be {\em conserved} along the
characteristics (solid blue line shown in Fig.~\ref{characFig}) and
this alone would give about $0.2\%$ shift in the acoustic peak
position by $z=0$.  However, as mentioned above, the correlation
function {\em grows} along the characteristics.  This growth is
governed by the divergence of the infall velocities, and, for large
$\eta$, it is this contribution which dominates.  Indeed, we have not
yet even included the linear amplification of the correlation
function, resulting from the right hand side in Eq.~(\ref{alongC}).

Including the divergence of infall velocities using
Eq.~(\ref{thetalin}), makes Eq.~(\ref{formsol}) for the two-point
function
\beqa
 1+\xi(r, \eta) &= &\Big( 1+\xi_0[r_0(r,\eta)] \Big) \nonumber \\
  & & \times \exp \Big[2 \int_0^\eta \xi_0[r_{\eta'}(r,\eta)] \, 
                       {\rm e}^{2\eta'} d\eta' \Big]. \nonumber \\ 
  & &  
\label{sol}
\eeqa
If the flow of characteristics caused by the nonlinear term in
Eq.~(\ref{paircon3}) is ignored, then $r\approx r_{\eta'} \approx
r_0$, and so
\beqa
 1+\xi(r) & \approx & \Big(1+\xi_0(r)\Big)\, 
            \exp \Big[ \xi_0(r)({\rm e}^{2\eta} -1) \Big] \nonumber \\
 &\approx & 1 +\xi_0(r)\, {\rm e}^{2\eta}.
\label{linrecover}
\eeqa
The final expression follows if the term in the exponential is small;
notice that it equals the linear perturbation theory expression for
$\xi$ at time $\eta$.

At first sight, the solution of Eq.~(\ref{sol}) appears to require
many evaluations of Eq.~(\ref{prop}).  However, the integral over
$\eta^{\prime}$ in the exponential piece of Eq.~(\ref{sol}) may be
transformed using the characteristic curve, whence
\beq 2e^{2\eta} = d(e^{2\eta}-1) = -\frac{3}{\bar{\xi}(r)}\frac{dr}{r}
\ .\eeq
Thus on performing this change of variables, the term in the
exponential of Eq.~(\ref{sol}) becomes
\beq 
 \Rightarrow\  3\int_{r}^{r_0} \frac{dr'}{r'}\,\frac{\xi_0(r')}{\bar\xi_0(r')}\ .
 \label{eta2scale_pre}
\eeq
However, on noting that $d\left[r^3\bar{\xi}(r)\right]/r^3 = 3
\xi(r)dr/r$, we find that this may be further simplified to
\beq 
\Rightarrow\
 \int_{r^3\bar\xi_0(r)}^{r_0^3\bar\xi_0(r_0)} \frac{dx}{x}\ .
 \label{eta2scale}
\eeq
Therefore, Eq.~(\ref{sol}) is really rather simple:
\beq
 1+\xi(r,\eta) = \Big( 1+\xi_0(r_0) \Big)\, 
                  \frac{r_0^3\,\bar\xi_0(r_0)}{r^3\,\bar\xi_0(r)}\ , 
 \label{simplerxi}
\eeq
and a single evaluation of Eq.~(\ref{prop}) gives $r_0(r,\eta)$, 
and hence the nonlinear value of $\xi(r)$.  


\subsection{Connections to previous work}

At late times ${\rm e}^{2\eta}\gg 1$.  Hence, on the large scales
where $\xi_0\ll 1$, Eq.~(\ref{linrecover}) implies that
$1+\xi(r)\approx \exp[\Xi_\eta(r)]$, where $\Xi_\eta(r) = {\rm
e}^{2\eta}\,\xi_0(r)$ is the linearly evolved correlation function.
This is precisely the relation between the correlation function of a
lognormal field and that of the underlying Gaussian field from which
it was derived.  Of course, this analysis has assumed that $r\approx
r_{\eta'} \approx r_0$; Figure~\ref{characFig} shows that this is
inappropriate at late times.  Nevertheless, it provides a nice
illustration of why the Lognormal has proved to be such a useful
approximation, and why the approximation breaks down~\footnote{See
\cite{ColesJones1991} for a discussion of how the Lognormal appears
from consideration of the continuity equation of $\delta$ and $v$,
Eq.~(\ref{cont}), itself.}.  Note that, both in linear theory and in
the Lognormal approximation for the nonlinear evolution, the position
of the acoustic peak does {\em not} shift~\footnote{The Lognormal
mapping is an example of a nonlinear transformation that does not
generate a shift in the acoustic peak, despite having
mode-coupling. The reason for this is that the mapping is local,
$\xi(r)$ is related to $\xi_{\rm linear}(r)$ at the same
scale. Gravitational instability is nonlocal and generates mode
coupling that generically leads to shifts,
see~\cite{CrocceScoccimarro2007} for more discussion along these
lines.}.

Our Eq.~(\ref{simplerxi}) has the flavor of an approach pioneered by
\cite{Hamiltonetal1991,NityanandaPadmanabhan1994}, who argued that
\beq
 1+\bar\xi(r,\eta) = (r_0/r)^3
\eeq
should provide a good approximation to nonlinear evolution.  
In effect, their approach sets 
\beq
 1+\xi(r,\eta) = \left(\frac{r_0}{r}\right)^3\, 
                 \frac{\partial\ln r_0}{\partial\ln r}.  
\eeq
If we set $1+\xi_0(r_0)\to 1$, then we have 
\beq
 1+\xi(r,\eta) = \left(\frac{r_0}{r}\right)^3
                 \frac{\bar\xi_0(r_0)}{\bar\xi_0(r)};
\eeq
our expression follows from inserting the linear velocities in the
characteristics---it is not an ansatz.  Note that this relation
changes if nonlinear velocities are used.


\begin{figure}
\begin{center}
{\includegraphics[width=0.5\textwidth]{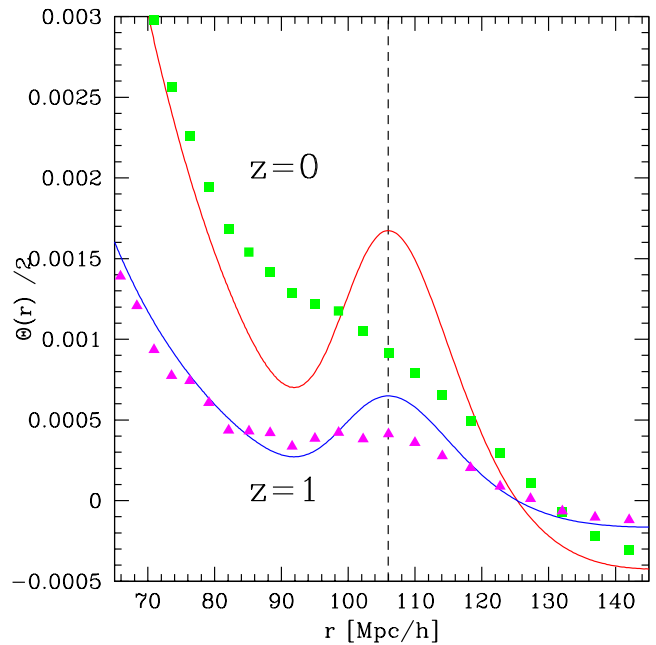}}
\caption{The divergence of pairwise infall (dark matter) velocities
$\Theta$ as a function of scale measured in numerical simulations at
$z=0$ (solid squares) and at $z=1$ (solid triangles).  In linear
perturbation theory, $\Theta/2$ is equal to the linear correlation
function -- and at $z=0$ and $z=1$ the linear models are represented
by solid red and blue lines, respectively. Notice that at $z=1$ the
acoustic peak is visible in $\Theta$, but by $z=0$ the nonlinear
effects have completely washed out any trace of it.}
\label{ThetaNB}
\end{center}
\end{figure}


\subsection{Inaccuracy of linear theory velocities}

In linear theory the divergence of infall velocities $\Theta(r,\eta)$
is, modulo a factor of two, given by the linear two-point function
itself (c.f. Eq~\ref{thetalin}) and this has a static (independent of
$\eta$) peak at the unperturbed position.  Hence, there is a
competition between $\Theta$, which prefers the peak to stay
unshifted, and the flow of characteristics, which induce a shift
towards smaller scales (Fig.~\ref{characFig}). A consequence of this
is that, {\em using linear velocities is expected to underestimate the
true peak shift}.  (The top left panel of Fig.~\ref{ximodel} shows
this explicitly, as we discuss later.)  Using Eq.~(\ref{simplerxi})
one obtains a shift of about $0.1\%$ at $z=0$, half of that due to the
flow of characteristics.

This underestimate results from the fact that, whilst the pairwise
infall velocity may be reasonably well described by linear theory on
large scales, its divergence deviates from linear theory more
strongly, due to the scale dependence of nonlinear corrections
\cite{Scoccimarro2004,CrocceScoccimarro2006b}.  This is graphically
illustrated in Fig.~\ref{ThetaNB}.  Although $\Theta\propto\xi$ in
linear theory, by $z=0$, nonlinear effects have washed out any sign of
an acoustic peak in $\Theta$!

In practice, a characteristic that probes scales slightly smaller than
the unperturbed acoustic peak will experience {\em more} growth of the
two-point function at late times.  This leads directly to an
enhancement that dominates over the effect of the flow of
characteristics, and results in a substantially enhanced shift over
the linear case (and as we will show this enhancement is about one
order of magnitude). In this sense the flow of characteristics only
gives a {\em lower} bound to the shift in the peak position due to
mode coupling.  Clearly, in order to proceed, we require a model for
the nonlinearity of the infall of pairwise velocities, and in
particular its divergence $\Theta$.


\subsection{Beyond linear theory velocities}

There are two types of nonlinear contributions to the pairwise infall
velocity.  This can be seen more clearly by rearranging
Eq.~(\ref{v12}) into the form,
\beq
\u_{12} = \frac{\langle (\delta_1+\delta_2)(\u_1-\u_2) \rangle
              + \langle \delta_1 \delta_2 (\u_1-\u_2) \rangle}{1+\xi}.
\label{u12}
\eeq
If we insert the standard perturbation theory (hereafter, PT)
expansions for $\delta$ and $u$ \cite{Bernardeauetal2002}, then we see
that the first term in the numerator is second order in
$\delta(\bx,\eta_0)$, and the second term is of third order, which in
linear theory averages to zero.  We can set the denominator equal to
unity, since $\xi$ is of order $10^{-3}$ on the scales of interest and
we are after much larger ($1-10\%$) effects. As mentioned in the
previous subsection, the effects from nonlinear mode-coupling on
$u_{12}(r,\eta)$ on these scales are negligible ($\sim1\%$), and hence
play almost no role in shaping the characteristic curves (which, as we
said, lead to shifts of only $\sim0.2\%$ in linear theory for dark
matter). They {\em do}, however, have a significant impact on the
source term in the right-hand-side of Eq.~(\ref{alongC}), which
dictates how fast the two-point function grows along the
characteristics.

Eq.~(\ref{u12}) thus leads to the following decomposition
\beq \Theta(r,\eta) \equiv \Theta_2(r,\eta)+ \Theta_3(r,\eta),
 \label{thetasplit}
\eeq
where the two terms on the right-hand-side are defined
$\Theta_2\equiv2\nabla\cdot \left<\delta_1\u_2\right>$ and
$\Theta_3\equiv 2\nabla\cdot \left<\delta_1\delta_2\u_1\right>$.
Considering the first term, on using the standard PT expansions for
the density and divergence of the velocity field
(\cite{Bernardeauetal2002} and see also footnote
\cite{velocityfield}), we find that $\Theta_2$ can be written
\beq
\Theta_2(r,\eta) = 2\,  \int P^{\delta\theta}\!(k,\eta)\, j_0(kr)\, d^3k\ , 
\label{theta2}
\eeq
and
\beq
P^{\delta\theta}\!(k,\eta) = {\rm e}^{2\eta}\, P^{\delta\theta}_0(k) 
                         + {\rm e}^{4\eta}\, P^{\delta\theta}_{\rm 1loop}(k)\ ,
\label{Pdethe}
\eeq
is the cross-power spectrum of the density and velocity divergence
expanded to fourth order in the standard PT.  The first term is the
usual one from linear theory $P^{\delta\theta}_0=P_0$, and
$P^{\delta\theta}_{\rm 1loop}$ is the `one-loop' correction to
$P^{\delta\theta}$ from PT.  The middle panel of Fig.~6 in
\cite{Scoccimarro2004} shows that this term describes rather well
(much better than for the density power spectrum) the deviations from
linear theory at large scales. Thus,
\beq \Theta_2(r,\eta) = \Theta_{2}^{0}(r,\eta) 
                      + \Theta_{2}^{\rm 1loop}(r,\eta).
\label{theta2break} 
\eeq

Considering the second term in Eq.~(\ref{thetasplit}), we find that
\beqa
 \Theta_3(r,\eta) &=& 2\, \nabla \cdot \langle \delta_1 \delta_2 \u_1 \rangle\ ,
  \nonumber \\ 
 &=& 2 \, {\rm e}^{4\eta} \int d^3k_1d^3k_2\,
    {\rm e}^{i \k_{12}\cdot \r}\ \frac{\k_{12}\cdot \k_2}{k_2^2}\,
    \nonumber \\
 & & \times \ B^{\delta\theta\delta}(\k_1,\k_2) \ ,
\label{theta3}
\eeqa
where $\r=\x_1-\x_2$, $\k_{12}=\k_1+\k_2$ and $B^{\delta\theta\delta}$
is the density--velocity divergence--density bispectrum: $\langle
\delta(\k_1) \theta(\k_2) \delta(\k_3) \rangle =
B^{\delta\theta\delta}(\k_1,\k_2)\, \delta_{\rm D}(\k_1+\k_2+\k_3)$.
Appendix~\ref{appen} provides explicit expressions for $\Theta_2^{\rm
1loop}(r)$ and $\Theta_3(r)$ expressed up to 1-Loop in the standard
PT, and written in terms of the initial power spectrum.

A substantially improved model for the nonlinear correlation function
$\xi$ results from including these nonlinear terms in
Eq.~(\ref{formsol}).  Before showing this explicitly, the next
subsection discusses how the effects of galaxy/halo biasing can be
included in our analysis.


\subsection{Extension to biased tracers}

The analysis above has been useful for understanding the motion of the
acoustic peak in the dark matter correlation function.  However, since
the observations will not measure the mass directly, but instead the
clustering of some set of biased tracers of the density field,
i.e. some sampling of the galaxy distribution, the method of
characteristic curve solutions will be more useful if we can extend it
to describe these biased tracers.  At first glance, it is not obvious
that this can be done, owing to the fact that halos, and the galaxies
that they host, are created and destroyed through merging, so their
comoving number density is not conserved. Thus one might na\"ively
conclude that any such approach based on continuity arguments must be
suspect. However, some thought shows that this problem is not
insurmountable.

Consider the motion of some halo today, its trajectory is the result
of the previous history of motions of its constituent particles.
Thus, for instance, one {\em may} speak of the motion of the center of
mass of the particles that make up the halo, at, say, the present
time.  In particular, one may also speak of the position and velocity
of its center of mass even at high redshifts when the halo itself does
not yet exist as a single virialized entity.  This was the point made
by \cite{Shethetal2001}; provided appropriate care is taken of how the
bias associated with these tracer particles evolves, the continuity
equation {\em can} indeed be used to relate $\xi$ to $v_{12}$.  The
argument above remains true if each halo is represented not by one but
by many tracer particles, and the number of tracer particles depends
on halo mass.  The positions of each of these tracers can be followed
back in time, so their number is conserved.  These tracers have some
effective bias factor at the time they are identified; provided one
accounts for the evolution of this bias, the continuity equation can be
used.  Since the argument above works for any set of tracers, it is as
valid for galaxies as for halos.  Note in particular that detailed
knowledge of the origin of the effective bias factor is unnecessary.
E.g. if two sets of tracers have the same abundance and bias factor at
one epoch, but one tracer populates a wide range of halo masses, and
the other two narrow but rather separate mass bins, the evolution of
the effective bias factor will be the same.

Fortunately, describing the evolution of the bias for `objects' that
are neither created nor destroyed is rather straightforward
\cite{NusserDavis1994,MoWhite1996,Fry1996,TegmarkPeebles1998,ShethTormen1999,Shethetal2001}:
For a set of tracer particles that are related to the underlying dark
matter through a linear, local, deterministic mapping, the time
evolution of their bias
($b(\eta)\equiv\delta^{\alpha}(\bx,\eta)/\delta(\bx,\eta)$ where
$\alpha$ represents either haloes or galaxies), can be written
\beq
 b(\eta)-1 = (b_i-1) {\rm e}^{-\eta}\ ,
 \label{bevol}
\eeq
where $b_i$ denotes the bias at the initial time $\eta=0$. Thus to
incorporate this bias model into our theoretical model, we must simply
make the following replacements:
\beq
 \xi_0 \rightarrow b_i^2\, \xi_0 \ ;\ \ \
 \Theta_2 \rightarrow  b(\eta)\, \Theta_2\ ; \ \ \
 \Theta_3 \rightarrow  b(\eta)^2\, \Theta_3\ ,
 \label{biased}
\eeq
in the expressions above.  Here we have used the standard assumption
that the velocity field of any set of biased
tracers is itself unbiased, and that $\Theta_3$ depends only 
quadratically on the density field, where we have neglected
sub-leading terms (see \cite{Fry1996}).

With these changes, Eq.~(\ref{prop}) for the characteristics becomes
\beq
 {\rm e}^{2\eta}-1+2(b_i-1) ({\rm e}^{\eta}-1) 
   = \int_{r}^{r_0} \, \frac{dr}{\bar\xi_0(r)}\ .
 \label{propbias}
\eeq
Fig.~\ref{characFig} shows solutions to this expression for tracers
that have bias factors of $b=1.4$ and $b=2$ at $z=0$.  It shows that
the flow of characteristics towards small scales is enhanced if $b>1$;
and this is as expected, because infall velocities are proportional to
the bias factor \cite{Shethetal2001}.

Our model for the nonlinear correlation function of biased tracers 
means that Eq.~(\ref{formsol}) becomes  
\begin{widetext}
\beq
1+\xi(r, \eta)= \Big( 1+b_i^2\, \xi_0[r_0(r,\eta)] \Big)  
  \times \exp \Big[ \int_0^\eta d\eta'  
  \Big(b(\eta')\,  \Theta_2[r_{\eta'}(r,\eta),\eta'] + 
  b(\eta')^2\,  \Theta_3[r_{\eta'}(r,\eta),\eta']\Big)\, \Big].
\label{formsolb}
\eeq
\end{widetext}
Note that the linear theory solution of this equation may be recovered
directly by setting: $\Theta_2=\Theta_2^{0}$; $\Theta_3=0$; and
$r_\eta' = r = r_0$ in the expression above. Whence,
\beq
 \xi(r)  \approx b(\eta)^2\, \xi_0(r)\, {\rm e}^{2\eta}\ ,
 \label{linrecoverb}
\eeq
and this is the generalization of Eq.~(\ref{linrecover}).


\begin{figure*}
\begin{center}
{\includegraphics[width=1.05\textwidth]{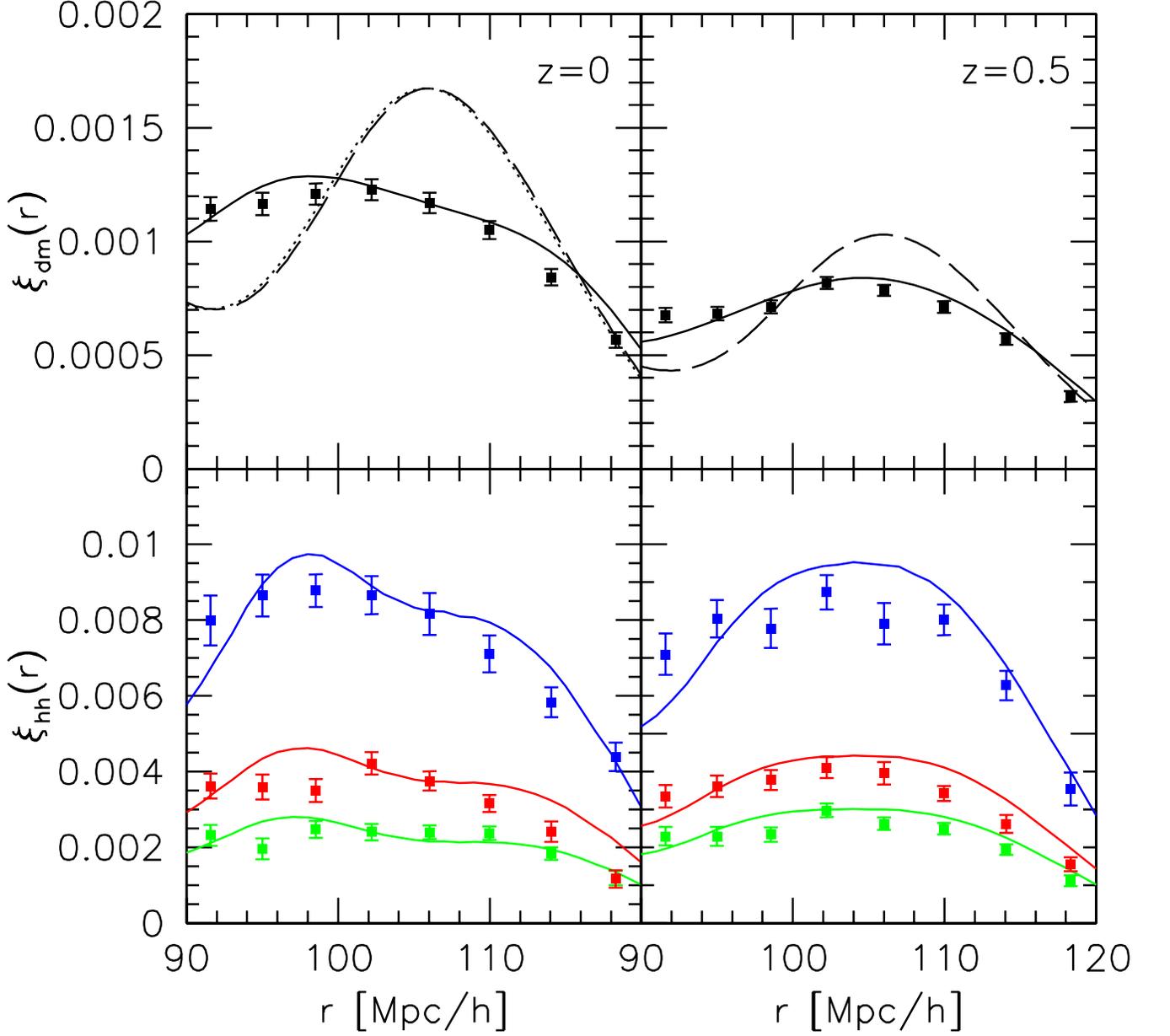}}
\caption{The real-space two-point correlation function for dark matter
(top) and halos (bottom) at $z=0$ (left) and $z=0.5$ (right).
Table~\ref{halocat} describes the three bins in halo mass; $\xi_{\rm
hh}$ is largest for the most massive halos.  The dashed lines in the
top panels show linear theory for the dark matter, solid lines are the
predictions of our model, Eq.~(\ref{formsolb}), and symbols show the
measurements.  Dotted line in the top left panel shows our model when
only linear theory velocities are used; it is almost indistinguishable
from simple linear theory, demonstrating that inclusion of the
nonlinear contributions to the (divergence of the) velocity field is
vital. }
\label{ximodel}
\end{center}
\end{figure*}


\subsection{Comparison with simulations}

Figure~\ref{ximodel} compares our model for the nonlinear correlation
function, Eq.~(\ref{formsolb}), with our measurements of (real-space)
$\xi$ for the dark matter (top) and halos (bottom) at $z=0$ (left) and
$z=0.5$ (right).  The halo measurements are the same as those
presented previously, except that now we only show scales which are
within $\sim 15\Mpc$ of the initial acoustic peak.

Our model for the dark matter, Eq.~(\ref{formsol}), matches the
measurements rather well; the solid lines are a substantial
improvement over linear theory (dashed).  Our model predicts that the
peak has shifted to $105 \Mpc$ by $z=0.5$, about a one percent effect;
this is in good agreement with a more rigorous calculation based on
RPT \cite{CrocceScoccimarro2007}.  By $z=0$ our model for $\xi_{\rm
dm}$ predicts that the peak has shifted to $98 \Mpc$, roughly an
$\sim8$ percent shift! This disagrees by over a factor of four with
the RPT calculation (apparent and physical shifts included); this
overshoot is not surprising, when given the fact that one-loop PT is
known to overestimate the nonlinear power spectrum by tens of percent
on small scales, even though the one-loop density-velocity divergence
power spectrum does well at reproducing the cross-power spectrum as
measured from numerical simulations at intermediate
scales~\cite{Scoccimarro2004}.

Turning now to the results for the dark matter halos, we see that
Eq.~(\ref{formsolb}) provides a very good description of the
measurements.  We emphasize that {\em there are no free parameters in
this model}. The only non-cosmological parameters in the model are the
bias factors and as discussed earlier, these are measured directly
from the simulations to make the predictions (see Section
\ref{thesignal} for our estimated values for the halo mass bins listed
in Table~\ref{halocat}).

When the bias factor is large, then the dominant nonlinear correction
comes from $\Theta_3$ because it scales as $b^2$.  For the dark
matter, the nonlinear correction coming from $\Theta_2^{\rm 1loop}$ is
the dominant one.  The figure shows that our model does not predict
any significant trend with halo mass, although this would likely
change if we were to include nonlinear bias (e.g. \cite{Smithetal2007}
suggest higher mass halos will show enhanced nonlinear effects).  Our
model requires knowledge of how these nonlinear bias terms evolve
(i.e., the analog of Eq.~\ref{bevol}): this evolution is given in
\cite{Scoccimarroetal2001,CooraySheth2002}.


\section{Conclusions}
\label{concl}

We have used analytic methods and a very large ensemble of numerical
simulations to study how the position of the baryonic acoustic peak in
the two-point correlation function, $\xi$, remnant of the tight
coupling between photons and baryons before recombination, is affected
by the clustering systematics: nonlinear mass evolution, bias and
redshift space distortions; and we have examined these effects as a
function of cosmological epoch and as a function of several trace
particle types -- i.e. halo samples picked to evolve with constant
comoving number density.

We have investigated a toy-model for the evolution of $\xi$ 
(Sec \ref{getreal}) that was simply a Gaussian bump plus a power-law 
and we showed that, if nonlinear evolution was manifest as a Gaussian
smoothing of the true $\xi$, then the acoustic scale was not well
recovered through simply measuring the local maximum -- and this we
described as {\em an apparent shift of the peak}. However, if there
was a change in the underlying broad band power then this we said
would lead to a {\em a physical motion of the peak}.

We presented results from our numerical simulations (Sec \ref{sims}). 
Our total simulated volume corresponded to $\sim~105~{\rm Gpc}^3h^{-3}$, 
approximately the same size volume that the proposed
 Stage IV, JDEM mission, ADEPT intends to survey\cite{Bennettetal2006}. 
Therefore our results and analysis are of direct relevance to that and 
similar missions. From these simulations we measured $\xi$ for the 
dark matter and haloes.  We found, at $z=0$ in both real and redshift 
space, that the true position and shape of the linear theory function 
did not match well that of the measured data -- there being an enhanced 
signal on scales smaller than the unperturbed peak scale.

We then performed a more careful analysis, and fitted the correlation
function data using the Gaussian smoothed linear theory model. This
provided a much improved fit. In all cases the inferred peak positions
from these models were shifted to smaller scales, with typical shifts
being of the order $\sim0.5-3.0~h^{-1}\,{\rm Mpc}$ -- including a
factor of $\sim2$ correction for the transfer function; the shifts
were enhanced for the the highest mass haloes/rarest objects and in
redshift space. However they were alleviated for higher redshifts. We
concluded that this was direct evidence for `apparent motion' of the
acoustic peak. We also noted that many of the recently proposed BAO
reconstruction methods do attempt to account for this apparent
shifting of the peak.

We then showed that the smoothed linear model was not a perfect fit to
the data, in particular for highly biased haloes and their galaxies
the fit was poor. Using the $\chi^2$ test we ruled out this model at
the 99\% significance. Furthermore, through inspection of the
residuals of the fitting we found $\sim 10-20\%$ excess of amplitude
on scales smaller than the unperturbed acoustic scale. We concluded
that this was supporting evidence for the hypothesis that non-linear
evolution was inducing a {\em physical} motion of the acoustic peak.
We characterized the physical shifts by finding the local maximum of
smooth polynomial fit to the data and subtracting from it the linear
model peak position. We found that these shifts were of the order
$\sim0.0-3.0~h^{-1}\,{\rm Mpc}$ for the samples considered.  
We noted that these -- which represent broad band tilting plus the 
more pernicious mode coupling effects -- are not accounted for 
in recently proposed schemes to correct the signal in the 
power-spectrum. 

In our analysis of the simulation data we also presented evidence 
that the simple Gaussian-based calculation for the variance
(Eq.~\ref{covar}) of $\xi$ that included the Poisson shot-noise
contribution provided a good description of the expected error on the
measured $\xi$ for haloes (Figs.~\ref{xirFit} and~\ref{xisFit}). In
detail the Gaussian error model was found to underestimate the
simulations for haloes at the present day. Adding non-Gaussian terms
from the trispectrum and bispectrum may improve this further.

For future BAO missions that aim to use the power spectrum of
clusters, the expected sample variance estimates that use the 
Gaussian plus Poisson model, will give reasonable estimates of the
variance. Directly extrapolating our analysis to make a statement
about the variance estimates for BAO galaxy surveys is complicated. 
Our analysis has dealt with the clustering of the halo centers and 
does not include the virial motions of particles/galaxies internal 
to the halo -- adding in this layer of reality may lead to increased 
variance. Furthermore, if galaxies appear only in haloes,
then using the galaxy number density estimate as the Poisson shot
noise error as is common practice, will underestimate the sample
variance when there is more than one galaxy per halo. We expect that
the effective number density of the haloes that host the galaxies in
the survey will be a better reflection of the errors. We shall reserve
this issue for future study.

We then presented an analytic model that was able to capture the main
observed affects from the non-linear evolution of the mass and
bias. The model was based upon a study of the gravitationally driven
mean streaming motions of particle pairs. These motions both smooth
out the initial peak, and, more importantly, shift it
(Fig.~\ref{characFig}).  In essence, our model simultaneously accounts
for {\em both} the smoothing and the shifting of the acoustic peak.
We first discussed the model in the context of the dark matter
(Eq.~\ref{formsol}), and then showed how it could be extended to
describe the nonlinear evolution of $\xi$ for biased tracers, such as
galaxies and clusters of galaxies as well (Eq.~\ref{formsolb}). For
the dark matter, our approach is less reliable than that of RPT (see
\cite{CrocceScoccimarro2007} for a discussion of this).  However, we
think it has substantial merit, owing to the fact that it permits a
simple description of how the shifting of the acoustic peak is
modified for biased tracer particles. It also allows us to see the
problem from a different perspective. One could combine the strength
of both methods, by replacing the modeling of the divergence of
pairwise velocities by its RPT description, for that one would need to
calculate the bispectrum contribution to $\Theta_3$.

The measured shifts in the acoustic peak position for the dark matter
and the biased tracers, are qualitatively consistent with the effects
of the clustering systematics on the power spectrum
\cite{Smithetal2007}. This owes to the fact that the power spectrum
and correlation functions are a Fourier transform pair. Thus small
scale damping and tilting of the linear power spectrum can lead to
both smoothing and tilting of the correlation function, and the
generation of the measured apparent and physical motion of the peak.

However, these recovered shift values appear substantially larger than
those currently quoted in the literature from other analytic arguments
\cite{Eisensteinetal2006a}.  One possible explanation for this is that
the divergence of the pairwise velocity field is substantially more
non-linear than the density field on these large scales
(Fig.~\ref{ThetaNB}). Had we simply used linear theory velocities in
our analytic model then we would have considerably underestimated the
measured shifts. Using perturbation theory was crucial
(Eqs.~\ref{thetasplit}--\ref{theta3}) for our model to get the close
agreement with the numerical measurements.

If unaccounted for, the percent level changes we have measured in the
acoustic scale will lead to biased determinations of cosmological
parameters (and see \cite{Anguloetal2007}). However, the agreement
between our model and the simulations suggests that, although such
pernicious shifts are present, it may be possible to construct
analytic tools that allow us to correct for them.  This is the subject
of ongoing work.


\vspace{0.5cm}

\acknowledgements

We would like to thank Gary Bernstein, Martin Crocce, Jacek Guzik,
Bhuvnesh Jain, Laura Marian, Nikhil Padmanabhan, Uros Seljak and
Martin White for useful discussions. We kindly thank Volker Springel
for making public his {\tt GADGET-2} code.  RES and RKS acknowledge
support from the National Science Foundation under Grant
No. 0520647. RS is partially supported by NSF AST-0607747 and NASA
NNG06GH21G. Lastly we thank G. Galilei for suggesting the title.


\appendix

\begin{widetext}

\section{The Divergence of Infall velocities in Perturbation Theory}
\label{appen}

This Appendix provides expressions for $\Theta_2^{\rm 1loop}(r)$ and
$\Theta_3(r)$ from the standard PT.


\subsection{$\Theta_2^{\rm 1loop}$ in the standard PT}

$\Theta_2^{\rm 1loop}$ is given by Eqs.~(\ref{theta2})
and~(\ref{Pdethe}), and is an integral over the 1-Loop contribution to
the velocity divergence-density power spectrum. In the standard
Perturbation Theory this can be written \cite{Scoccimarro2004}:
\beq
 P^{\delta\theta}_{\rm 1loop}(k) = 
  2 \int F_2(\k-\q,\q)\, G_2(\k-\q,\q)\, P_0(|\k-\q|)\, P_0(q)\, d^3q 
  + 3\, P_0(k) \int [\hat{F}_3(k,q) + \hat{G}_3(k,q)]\, P_0(q)\, d^3q,
 \label{Pdetheloop}
\eeq
where the functions $F_2(\k,\q)$ and $G_2(\k,\q)$ are the second
order, symmetric, density and velocity divergence kernels from PT
\cite{Bernardeauetal2002}. These are written:
\beqa
F_2(\k,\q) &=& \frac{5}{7} +\frac{1}{2}\, \mu_{k,q} \, 
\Big(\frac{k}{q}+\frac{q}{k}\Big) +
\frac{2}{7}\, \mu_{k,q}^2 \ ;\\
& & \nonumber \\
G_2(\k,\q) &=& \frac{3}{7} +\frac{1}{2}\, \mu_{k,q} \, 
\Big(\frac{k}{q}+\frac{q}{k}\Big) +
\frac{4}{7}\, \mu_{k,q}^2 \ ,
\eeqa
where $\mu_{k,q}\equiv \k\cdot\q/|k||q|$. The functions
$\hat{F}_3(k,q)$ and $\hat{G}_3(k,q)$ are the angle averages of the
third order PT density and velocity kernels. These may be written:
\beqa
\hat{F}_3(k,q)&=& \int \frac{d\hat{q}}{4\pi} F_3(\k,\q,-\q)=
\frac{1}{24}
\left[ 
\frac{6k^6-79k^4q^2+50k^2q^4-21q^6}{63k^2q^4}+
\frac{(q^2-k^2)^3 (7q^2+2k^2)}{42k^3q^5}\ \ln \Big|\frac{k+q}{k-q}\Big| 
\right] \ ;
\\ \nonumber \\
\hat{G}_3(k,q)&=&\int \frac{d\hat{q}}{4\pi} G_3(\k,\q,-\q)= 
\frac{1}{24}
\left[ 
\frac{6k^6-41k^4q^2+2k^2q^4-3q^6}{21k^2q^4}  +
\frac{(q^2-k^2)^3 (q^2+2k^2)}{14k^3q^5}\ \ln \Big|\frac{k+q}{k-q}\Big| 
\right]\ .
\eeqa
%


\subsection{$\Theta_3(r)$ in the standard PT}

$\Theta_3(r)$ is related to the density-velocity divergence-density
bispectrum through two Fourier transforms (Eq.~\ref{theta3}). In the
standard PT this bispectrum is:
\beq
 B^{\delta\theta\delta}(\k_1,\k_2,\k_3) = 2 F_2(\k_2,\k_3) P_0(k_2)P_0(k_3) 
 + 2 G_2(\k_1,\k_3) P_0(k_1)P_0(k_3) + 2 F_2(\k_1,\k_2) P_0(k_1)P_0(k_2) \ .
\label{crossbisp}
\eeq
In order to proceed we require some further pieces of
information. Firstly, the closure relation for $k$-modes gives us
$\k_3=-\k_1-\k_2$. Secondly, statistical homogeneity and isotropy
means that the bispectrum can be written as a function of three
variables: the length of two sides of a triangle and the angles
between them, i.e. we should at the end of our calculation be able to
write $B^{\delta\theta\delta}(\k_1,\k_2,\k_3)\equiv
B^{\delta\theta\delta}(k_1,k_2,\theta_{12})$. Thirdly, the addition
theorem for spherical harmonics allows us to re-write the angles
between any two vectors in terms of their own angles in some arbitrary
Cartesian system:
\beq \cos\theta_{12} = \cos\theta_1\cos\theta_2+\sin\theta_1
\sin\theta_2\cos(\phi_1-\phi_2) \ ,\eeq 
where the angle between the two vectors $\k_1~\{k_1,\theta_1,\phi_1\}$
and $\k_2~\{k_2,\theta_2,\phi_2\}$ is $\theta_{12}$. Some lengthy
algebra then leads us to the following expression for $\Theta_3(r)$:
\beqa
\Theta_3(r) &=& 2 \Bigg[ \int d^3k\, P_0(k)\, j_1(kr)\, k \int d^3q\, P_0(q) \Big[ 
\Big(\frac{q}{k} \tvk \cdot \tvq +1\Big) \frac{2G_2(\k,\q)}{|\k+\q|^2}-\frac{1}{3}
\Big(\frac{1}{q^2}+\frac{1}{k^2}\Big)\Big] + \frac{34}{21} \Psi_0^0(r) \Psi_1^{-1}(r) 
\nonumber \\ & & -\frac{2}{3}\Big[\Psi_2^0(r)\Psi_1^{-1}(r)+\Psi_2^{-2}(r)\Psi_1^{1}(r)\Big]
+\frac{1}{3}\Big[\Psi_0^0(r)\Psi_1^{-1}(r)+\Psi_0^{-2}(r)\Psi_1^{1}(r)\Big]+
\frac{8}{35}\Psi_2^0(r)\Psi_3^{-1}(r) \nonumber \\ & & 
-\frac{16}{105}\Psi_2^{0}(r)\Psi_1^{-1}(r) \Bigg]\ ,
\label{Theta3PT}
\eeqa
where we have introduced the useful auxiliary function
\beq
 \Psi_{\ell}^m(r) = \int d^3 q \, P_0(q) \, j_{\ell}(qr)\, q^m\ .
\eeq
\end{widetext}




\begin{thebibliography}{50}
\bibitem[\protect\citeauthoryear{}{}]{velocityfield} 
{In solving equations like $\nabla \cdot
{\v(\br,\eta)}=\theta(\br,\eta)$, we assume that the velocity field is
curl free, whence $\v(\bx,\eta)=\nabla\psi(\bx,\eta)$ and thus $\theta
= \del^2 \psi(r,\eta)$.  This may be solved most simply in Fourier
space: $\v(\bk)=i\bk \theta(\bk)/k^2$. \label{velocityfield}}
\end{thebibliography}
\end{document}